\documentclass{aa}
\input {epsf}
\usepackage{graphicx}
\input psfig.sty

\begin{document}

\newcommand{\mdotr}{$\dot{M}(r)$}
\newcommand{\mdot}{$\dot{M}$}

\title{XMM-Newton observation of M87}
\subtitle{I. Single-phase temperature structure of intracluster medium}

\offprints{K. Matsushita}
\date{Received Jun 17, 2001 ; accepted Jan 15, 2002}
\author{Kyoko  Matsushita\inst{1} \and
Elena  Belsole\inst{2}\and Alexis Finoguenov\inst{1}  \and Hans  B\"ohringer\inst{1} }
\institute{Max-Planck-Institut f\"ur Extraterrestrial Physik, D-85748 Garching, Germany,
\and Service d'Astrophysique, CEA Saclay, L'Orme des Merisiers  B\^at 709.
, F-91191 Gif-sur-Yvette Cedex, France.}

\abstract{
We report the results of a detailed analysis of the  temperature structure  of the X-ray 
emitting plasma halo of M~87, the cD galaxy of the Virgo Cluster.
 Using  the MEKAL model, 
the data provide strong indications that the intracluster medium 
has a single phase structure 
locally, except the regions associated to the radio structures.
The deprojected spectrum at each radius is well fitted by a single temperature 
MEKAL model, except for the very central region ($<$  2 arcmin) which seems to be
affected by the jet and radio lobe structure.
The temperature of the intracluster plasma is 1 keV at the center and gradually increases 
to 2.5 keV at 80 kpc.
 We have also fitted spectra using the APEC code. Although the large changes of the strength
of K$\alpha$ lines causes a  discrepancy between the Fe-L and Fe-K lines
for the APEC results,
the overall temperature structure  has not changed.
There is no sign of excess absorption in the spectral data.
The single-phase nature of the intracluster medium is in conflict with the standard cooling
flow model which is based on a multi-phase temperature structure.
In addition, the signature of gas cooling below 0.8 keV to zero temperature is not observed
as expected for a cooling flow.
The gravitational mass profile derived from the temperature and density distribution
of the intracluster gas shows two distinct contributions that can be assigned to the
gravitational potential of the cD galaxy and the cluster. The central temperature of
the intracluster medium agrees well with the potential depth and the velocity dispersion of
the cD galaxy. The latter result implies that the central region of the intracluster medium
is equivalent to a virialized interstellar medium in M 87.
}
\maketitle

\section{Introduction}

In the  cores of many clusters of galaxies, X-ray imaging data show a highly peaked surface 
brightness profile (e.g. Fabian et al. 1981).
The radiative cooling time in these regions is much less than a Hubble time. 
Without a heating process, the gas cools to low temperature and results in a ``cooling flow''
(Fabian 1994 for a review).
The mass flow rate, \mdot, that is deduced in the standard cooling flow model,
is approximately proportional to the radius. This implies that matter is
deposited throughout the entire cooling flow region.
It also implies that the gas in the cooling flow zone is ``multi-phase'' on scales small enough
that the inhomogeneities have escaped the observation so far.
 
ASCA and ROSAT observations confirmed the existence of cooler gas in the cores of cooling flow
clusters as expected (e.g. Allen \& Fabian 1994, Ikebe et al. 1999,
Ikebe 2001).
The presence of an  intrinsic absorber in excess of the galactic absorption, as 
it might be expected to arise from the accumulation of cold gas in the cooling flows,
was also inferred from the analysis of the ASCA spectra (e.g. Allen 2000; Allen et al. 2001).
In fact, the fitting of the ``multi-phase'' cooling flow model to the observed spectra  including 
temperature phases that cool below the X-ray emitting temperature regime, 
requires the inclusion of excess absorption to  satisfactorily reproduce the observed spectra.
It has also been argued, that inclusion of excess absorption leads to values of the mass flow
rate as obtained from imaging data, $\dot{M}_{\rm{I}}$  in good agreement with those obtained from
spectral analysis, $\dot{M}_{\rm{S}}$ (Allen 2000).

However, without  assuming excess absorption, the values of $\dot{M}_{\rm{S}}$ are systematically lower than
$\dot{M}_{\rm{I}}$ (e.g. Ikebe et al. 1999; Makishima et al. 2001).
Instead of the cooling flow model, a two temperature spectral model can well fit ASCA
spectra of the cooling flow clusters.
The temperatures of the softer component are about half of the hotter component (Ikebe 2001). 
The central temperatures of some cD galaxies obtained with the ROSAT PSPC are close to those obtained
from normal elliptical galaxies with the same stellar velocity dispersion (Matsushita 2001).
Since cooling flow clusters always possess central dominant galaxies, 
the cool components may reflect the potential structure, rather than cooling gas (Ikebe et al. 1999,
Makishima et al. 2001, Ikebe 2001).

Recently, it was discovered with the RGS instrument onboard XMM-Newton that 
there is little X-ray emission from a component with a temperature
below a certain lower cutoff value that differs from object to object (e.g. Tamura et al. 2001; Kaastra et al. 2001).
However, with the RGS one observes only the very central part of the cluster.

M~87 is the cD galaxy of the nearest rich cluster of galaxies, the Virgo Cluster.
It is very luminous in X-rays, and  is suggested to have a ``cooling flow'' with 
\mdot~of about  10 $M_\odot\rm{yr^{-1}}$ (Stewart et al. 1984;  Fabian et al. 1984).
M~87 also hosts a central active galactic nuclei (AGN).
The radio emission is complex and there are two strong  lobe structures (e.g. B\"ohringer et al. 1995).
An enhancement of the X-ray emission around the lobes  was discovered with 
the EINSTEIN HRI (Feigelson et al. 1987) and ROSAT PSPC (B\"ohringer et al. 1995)

In  addition to the XMM RGS results mentioned above,
the XMM-Newton observatory offers the possibility to
perform a very detailed spectral and spatial analysis of the cooling flow area.
This is in particular important since the RGS spectrum taken at the very central part of
the intracluster medium (ICM) may be strongly affected by resonance line scattering and
thus the interpretation of the results requires the combined analysis of spectra taken
at different radii from the center.
Early results of the XMM observation of M 87 have  already been published (B\"ohringer et al. 2001; 
Belsole et al. 2001).
The projected spectrum is fitted with a single temperature model (B\"ohringer et al. 2001).
The abundance drop at the central region is interpreted as resonance scattering.
The detailed spectral analysis on the jets and  the radio lobe structure is shown in Belsole et al. (2001).

In this paper, we report on the more detailed temperature structure
using a deprojecting analysis,
and apply a cooling flow model to the spectral analysis.
We employ a distance of M~87 of 17 Mpc.
We adopt for the solar iron abundance the `photospheric'
value, Fe/H $=4.68\times 10^{-5}$ by number (Anders \& Grevesse 1989).
Unless otherwise specified,  we use 90\% confidence error regions.

\section{Observation and data preparation}
M87 was observed with XMM-Newton on June 19th, 2000.
The thin filter is used for  both detectors.

\subsection{EPN}

\begin{figure}[]
\resizebox{\hsize}{!}{\includegraphics{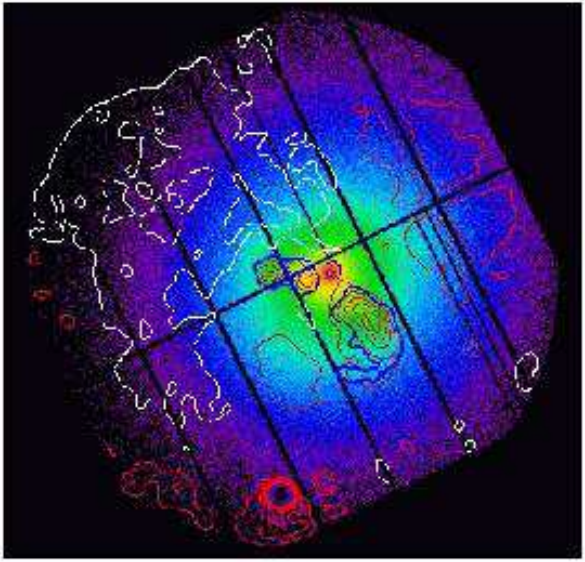}}
\caption{
EPN image of M87 in the energy range of 0.2 to 10.0 keV.
The effect of the out-of-time events is corrected.
The overlayed contours correspond to the image divided by a azimuthally symmetric 
with the same radial profile, smoothed with 20$''$ Gaussian.
The contour levels are 0.8, 0.9 (white), 1.1, 1.2, 1.3, 1.4, and 1.5 (red).
The regions surrounded by the blue contours are excluded from our spectral analysis as described in \S3.3.
}
\label{raw_spectra}
\end{figure}

Figure 1 shows the X-ray image of M87  obtained with the EPN.
The contribution from the out-of-time events is subtracted following the standard analysis package (SAS).
The image  divided by an azimuthally symmetric image with the same radial profile
is overlayed in Figure  1.
The X-ray halo is very smooth and almost spherically symmetric within 20\%,
except for the two inner radio lobe regions and the first knot in the jet.
A luminous point source  near the south edge of the  detector is excised in the further analysis.

We have selected single pattern events.
The background spectrum was calculated for all spectra
by integrating the Lockman-Hole data (observed in the Revolution
70,71,73,81 for the EPN) in the same detector regions.
In the Lockman-Hole field, sources with a  luminosity higher than that of
 the excluded sources from the M~87 region are excised.
The flare events in the background data were screened in the following way.
We made two count rate histograms of bin width of 200s,  
for the corner region in the whole energy band,
and  for  the whole detector for above 10 keV.
Then, we  fitted  each histogram with a gaussian, and selected the time 
within 3 $\sigma$ of the mean for both histograms.
The total exposure of the background data is 95ks.
We have also screened the M87 data in the same way.
The  exposure times of the EPN are 30ks, 28ks, 27ks, 30ks for the 
Chips 1-3, 4-6, 7-9, and 10-12, respectively.

The spectra of each CCD chip
are  accumulated within a  ring, centered on M~87.
Then, we have subtracted the background and corrected the effect of the vignetting and exposure,
and  summed the spectra within annular regions.
The contribution of out-of-time events is subtracted using SAS.


\subsection{EMOS}

We have selected events with pattern between 0 to 12. 
The background spectra are accumulated from the Lockman-Hole data
observed in  Revolution 70.
The background flare events are screened in the same way for the EPN.
The exposure times of the background and data are 33 ks and 40 ks, respectively.
Out-of-time events are not subtracted since the contribution is negligible.

\subsection{Spectral Deprojection}

Deprojected spectra are calculated by subtracting the contribution from the
outer shell regions for all spectral components, assuming the ICM is spherically symmetric 
 as done by Nulsen and B\"ohringer (1995).
Within a each shell, the spectrum per unit volume is assumed to be the same.
In order to calculate the 
 deprojected spectrum of the outermost ($N^{th}$) shell, we subtracted
 the emission of the outside of the $N^{th}$ shell  using the brightness
profile derived from the ROSAT All Sky Survey (B\"ohringer et al. 1994) 
 assuming that  spectral shape outside the $N^{th}$ shell is the same with that
of the $N^{th}$ shell.
The deprojected spectrum of the $i^{th}$ shell is then calculated by subtracting
 the  contributions from $i+1^{th}$ to $N^{th}$ shell and from outside of the
$N^{th}$ shell from the annular spectrum of the corresponding radius.

\section{Spatial distribution of temperature components}

\subsection{Radial temperature profile }

\begin{figure}[]
\resizebox{\hsize}{!}{\includegraphics{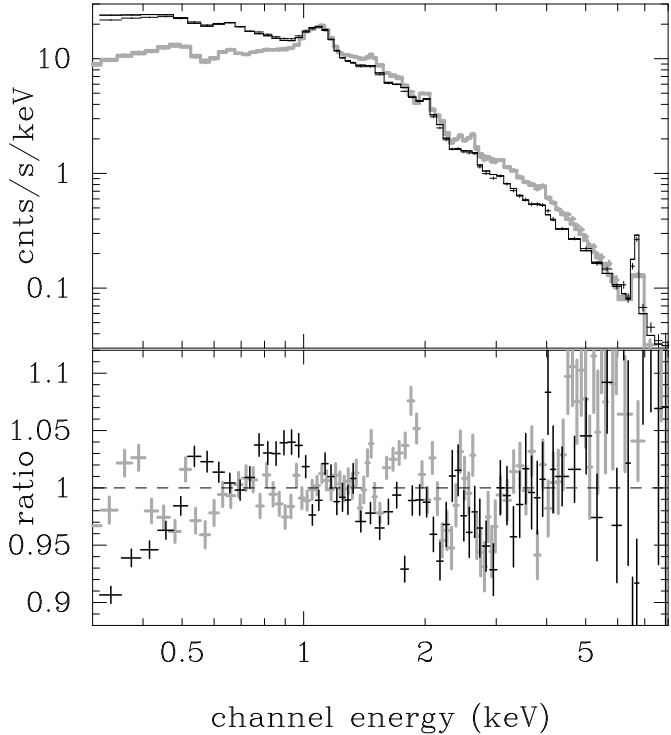}}
\caption{ Annular spectra of the EPN (black) and 
EMOS (EMOS1+EMOS2; gray) for $r$=5.6-8$'$,
fitted with a same single temperature MEKAL model.
Bottom panels show data-to-model ratios.
}
\end{figure}

We employ epn\_fs20\_sY9\_thin.rmf from March 2001, m1\_thin1v9q19t5r5\_all\_15.rsp
from June 2001 for the EPN and the EMOS response files, respectively.
The spectral analysis uses the XSPEC\_v11.0.1 package.

Figure 2 shows annular spectra for the radial range  $r$=5.6-8$'$ of the EPN and the EMOS,
fitted with a same single temperature MEKAL model (Mewe \& Gronenschild 1985, Mewe et al. 1986,
Kaastra 1992, Liedahl et al. 1995) but different normalization.
Here, $r$ is the the projected  radius.
The EMOS gives about 10\% higher normalizations than the EPN.
The response matrices of the  EPN and EMOS are mostly consistent.
Below 0.5 keV, EPN data is lower than the EMOS data, and there are 
small discrepancies around the tail of the Fe-L lines at 0.8-1.0 keV and around
the instrumental Si edge at 2 keV.
Since the EMOS gives a hydrogen column density, $N_{\rm{H}}$, consistent with that of ROSAT (c.f. \S6),
we fitted to the EPN and EMOS data in the spectral  range of 0.5 to 10 keV and 0.3 to 10 keV, respectively.
For $R>8'$, the energy range between 7.5 and 8.5 keV of the EPN spectra is ignored,
 because of strong instrumental emission lines.

We have fitted the spectra  for the region with radius  $R>0.5'$ 
with a MEKAL model with photoelectric absorption.
Here, $R$ is the three dimensional radius.
For the spectra within 0.5$'$,  a power-law component, with the
same absorption, $N_{\rm{H}}$, is added.
When fitting the spectra  between 0.12$'$ and 0.5$'$, we
fixed the index of the power-law component to the best-fit value obtained 
from the spectrum within 0.12$'$ and normalized it using the Point Spread Function.
We grouped several elements 
and constrained the elements in each group to have a common abundance.
The first group is C, N, and O, and the second group is Fe and Ni.
The other elements are determined separately.
We also fitted the spectra with a sum of two MEKAL models,
where the two components were constrained to have  common abundances.
%
\begin{table*}
\begin{tabular}[t]{rcccrrccrccrccr}
\hline
 & &\multicolumn{3}{l}{1T MEKAL} &\multicolumn{5}{l}{2T MEKAL}  \\ 
\hline
   R    & &   $kT$       &$N_H$  &$\chi^2/\mu^a$ &   $kT1$       & $kT2$
 & $N_H$  &$\chi^2/\mu^a$ & EM2/EM1$^b$  \\
(arcmin)& &  (keV)  &\tiny{($10^{20}\rm{cm^{-2}}$)}      & &(keV)  & (keV) &\tiny{($10^{20}\rm{cm^{-2}}$)} \\
\hline
0.00-0.12&EPN& 0.71$^{+0.09}_{-0.07}$ & 2.0 (fix) & 58/52 \\
0.12-0.25&EPN& 1.18$^{+0.08}_{-0.08}$ & 0.9$^{+1.9}_{-0.9}$ & 63/54 & \\
0.25-0.35&EPN& 1.22$^{+0.07}_{-0.07}$ & 2.8$^{+2.4}_{-2.1}$ & 74/54 & \\
0.35-0.50&EPN& 1.34$^{+0.07}_{-0.06}$ & 5.1$^{+2.1}_{-1.9}$ & 72/54 & 1.74$^{+0.19}_{-0.17}$ & 0.87$^{+0.13}_{-0.12}$ & 1.8$^{+2.4}_{-1.8}$ & 54/52 & 0.208$^{+0.133}_{-0.080}$\\
0.50-0.70&EPN& 1.41$^{+0.03}_{-0.03}$ & 5.1$^{+1.3}_{-1.1}$ & 90/54 & 1.72$^{+0.16}_{-0.10}$ & 0.94$^{+0.20}_{-0.12}$ & 2.4$^{+1.6}_{-1.7}$ & 57/52 & 0.156$^{+0.059}_{-0.043}$\\
0.70-1.00&EPN& 1.47$^{+0.03}_{-0.05}$ & 4.1$^{+0.9}_{-1.8}$ & 49/54 & 1.69$^{+0.13}_{-0.15}$ & 0.87$^{+0.31}_{-0.11}$ & 2.0$^{+2.4}_{-2.0}$ & 40/52 & 0.105$^{+0.065}_{-0.038}$\\
1.00-1.40&EPN& 1.43$^{+0.04}_{-0.04}$ & 4.9$^{+1.3}_{-1.5}$ & 40/54 & 1.54$^{+0.15}_{-0.05}$ & 0.89$^{+0.26}_{-0.19}$ & 3.8$^{+1.5}_{-2.0}$ & 33/52 & 0.090$^{+0.053}_{-0.051}$\\
1.40-2.00&EPN& 1.46$^{+0.03}_{-0.03}$ & 5.0$^{+1.1}_{-1.2}$ & 93/54 & 1.77$^{+0.12}_{-0.10}$ & 0.91$^{+0.12}_{-0.07}$ & 1.9$^{+1.4}_{-1.7}$ & 59/52 & 0.127$^{+0.044}_{-0.033}$\\
2.00-2.80&EPN& 1.78$^{+0.08}_{-0.05}$ & 2.6$^{+1.0}_{-1.0}$ & 42/54 & 1.95$^{+0.41}_{-1.07}$ & 1.31$^{+0.38}_{-0.39}$ & 2.2$^{+0.6}_{-1.2}$ & 37/52 & 0.025$^{+0.020}_{-0.020}$\\
2.80-4.00&EPN& 1.81$^{+0.09}_{-0.10}$ & 3.4$^{+1.2}_{-1.0}$ & 100/54 & 1.98$^{+0.09}_{-0.09}$ & 0.97$^{+0.19}_{-0.15}$ & 2.2$^{+0.9}_{-0.9}$ & 72/52 & 0.052$^{+0.017}_{-0.015}$\\
4.00-5.60&EPN& 2.06$^{+0.08}_{-0.06}$ & 2.0$^{+0.5}_{-0.4}$ & 120/53 & 2.19$^{+0.10}_{-0.07}$ & 0.97$^{+0.18}_{-0.16}$ & 1.5$^{+0.3}_{-0.9}$ & 93/52 & 0.043$^{+0.014}_{-0.012}$\\
5.60-8.00&EPN& 2.32$^{+0.08}_{-0.08}$ & 1.5$^{+0.6}_{-0.7}$ & 54/54 &
 2.37$^{+0.09}_{-0.08}$ & 1.0 (fix) & 1.3$^{+0.6}_{-0.6}$ & 43/56 & 0.023$^{+0.012}_{-0.012}$\\
8.00-11.30&EPN& 2.52$^{+0.07}_{-0.06}$ & 1.5$^{+0.2}_{-0.5}$ & 87/57 &
 2.59$^{+0.08}_{-0.08}$ & 1.0 (fix) &  1.4$^{+0.5}_{-0.5}$ & 77/56  & 0.022$^{+0.012}_{-0.012}$\\
11.30-13.50&EPN &2.43$^{+0.10}_{-0.09}$ & 1.3$^{+0.4}_{-0.7}$ & 114/57 &
 2.50$^{+0.14}_{-0.09}$ & 1.0 (fix) &  1.2$^{+0.4}_{-0.7}$ & 108/56  & 0.019$^{+0.015}_{-0.012}$\\
13.50-16.00&EPN& 2.55$^{+0.12}_{-0.12}$ & 0.6$^{+0.6}_{-0.6}$ & 68/57 & &&&  & 0.000$^{+0.019}_{-0.000}$\\
\hline
0.00-0.12 &EMOS&1.15$^{+0.24}_{-0.31}$ &  6.2$^{+1.1}_{-1.6}$ & 120.4/120\\
0.12-0.25&EMOS& 0.91$^{+0.13}_{-0.03}$ &  3.2$^{+0.8}_{-0.7}$ & 141.1/120\\
0.25-0.35&EMOS& 1.14$^{+0.05}_{-0.07}$ &  4.1$^{+1.4}_{-1.0}$ &
 110.9/120 &\\ 
0.35-0.50&EMOS& 1.44$^{+0.04}_{-0.06}$ &  3.5$^{+1.4}_{-1.2}$ &  124.5/116  &1.93$^{+0.35}_{-0.20}$ & 1.09$^{+0.17}_{-0.13}$ &  0.9$^{+1.4}_{-0.9}$ &    98.1/114& 0.172$^{+0.066}_{-0.048}$ \\
0.50-0.70&EMOS &1.37$^{+0.03}_{-0.04}$ &  5.0$^{+1.6}_{-0.8}$ & 156.6/116   &1.73$^{+0.15}_{-0.11}$ & 1.01$^{+0.10}_{-0.10}$ &  2.8$^{+1.2}_{-1.1}$ &    107.7/114& 0.184$^{+0.043}_{-0.036}$ \\
0.70-1.00&EMOS &1.52$^{+0.04}_{-0.03}$ &  3.8$^{+0.7}_{-1.1}$ &  113.3/116 &1.69$^{+0.09}_{-0.08}$ & 0.90$^{+0.21}_{-0.12}$ &  2.3$^{+1.0}_{-1.0}$ &   86.1/114& 0.083$^{+0.025}_{-0.022}$ \\
1.00-1.40&EMOS& 1.55$^{+0.05}_{-0.04}$ &  3.3$^{+0.8}_{-1.0}$ & 130.3/116 & 1.70$^{+0.08}_{-0.07}$ & 0.92$^{+0.20}_{-0.14}$ &  2.0$^{+1.0}_{-0.9}$ &  110.0/114  & 0.058$^{+0.021}_{-0.019}$ \\
1.40-2.00 &EMOS&1.38$^{+0.02}_{-0.02}$ &  4.7$^{+0.7}_{-0.9}$ &  145.3/116  &1.58$^{+0.07}_{-0.03}$ & 0.91$^{+0.10}_{-0.08}$ &  3.0$^{+0.9}_{-0.9}$ & 100.0/114  & 0.129$^{+0.024}_{-0.027}$ \\
2.00-2.80&EMOS& 1.88$^{+0.05}_{-0.05}$ &  2.0$^{+0.5}_{-0.5}$ &  146.4/116  &1.98$^{+0.08}_{-0.08}$ & 1.00$^{+0.19}_{-0.14}$ &  1.6$^{+0.3}_{-0.3}$ & 120.0/114  & 0.029$^{+0.009}_{-0.010}$ \\
2.80-4.00&EMOS& 1.83$^{+0.05}_{-0.04}$ &  2.9$^{+0.4}_{-0.4}$ & 165.4/116  &2.09$^{+0.23}_{-0.09}$ & 1.23$^{+0.22}_{-0.12}$ &  2.1$^{+0.5}_{-0.5}$ &  108.6/114  & 0.042$^{+0.010}_{-0.010}$ \\
4.00-5.60&EMOS& 2.12$^{+0.06}_{-0.05}$ &  0.8$^{+0.4}_{-0.4}$ & 140.3/116   &2.23$^{+0.10}_{-0.07}$ & 1.10$^{+0.24}_{-0.20}$ &  0.6$^{+0.4}_{-0.4}$ &    112.6/114& 0.017$^{+0.009}_{-0.009}$ \\
5.60-8.00&EMOS& 2.38$^{+0.06}_{-0.06}$ &  1.5$^{+0.4}_{-0.3}$ &  124.3/116 &   &  & &  & 0.002$^{+0.008}_{-0.002}$ \\
8.00-11.30 &EMOS&2.50$^{+0.07}_{-0.07}$ &  2.4$^{+0.3}_{-0.4}$ &  148.7/116 &  &  & &   & 0.002$^{+0.008}_{-0.002}$  \\
11.30-13.50&EMOS& 2.70$^{+0.05}_{-0.05}$ &  1.1$^{+0.2}_{-0.2}$ & 236.8/116 &  & & & & 0.000$^{+0.003}_{-0.000}$\\
\hline
\end{tabular}
a: degrees of freedom\\
b: Ratio of emission measure of the cooler and hotter component, when the temperature of the former is fixed to 1 keV.\\
\caption{Result of spectrum fitting of the deprojected spectra (whole region)}
\end{table*}

We have summarized the results in Table 1.

\begin{figure}[]
\resizebox{\hsize}{!}{\includegraphics{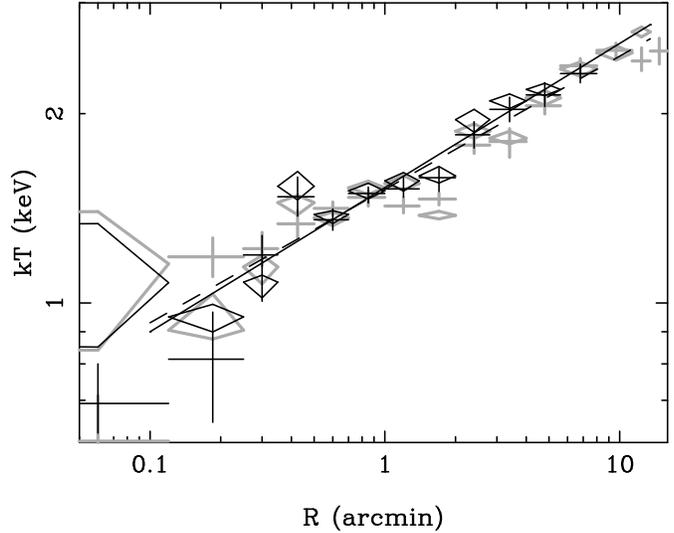}}
\caption{Deprojected radial profile of the temperature for the EPN (gray crosses) and the
EMOS (gray diamonds), obtained using the single temperature MEKAL model.
We also plotted the same result but excluding the jet and radio lobe
 regions (Figure1; \S3.3)
for the EPN (black crosses) and the EMOS (black diamonds).
The solid line and the dashed line corresponds to the best-fit power-law relation (excluding the
lobe regions) for the EMOS and EPN, respectively.
}
\end{figure}

Figures 3 and 4 show  the radial profile of the temperature obtained by the single temperature model,
and the two temperature model, respectively.
The temperatures derived from the EPN and EMOS are mostly consistent.

 Within $R<6'$ and $R<13.5'$ of the EMOS and EPN respectively,
 the two temperature model gives significantly lower $\chi^2$ values than the single temperature model.
The temperature of the cooler component is almost constant at $\sim 1$ keV,
which is the same as the central temperature obtained by the single temperature model fit.
That of the hotter component show a nearly  constant value at $R<2'$, whereas
for $R>2'$ it increases gradually.
In Figure 5, we have plotted the ratio of the emission measure of the cool to the hot component,
assuming the temperature of the former to be 1.0 keV.
  The fraction of the cool component is 20\% at $R$=$0.5'$ and drops
 below 1 \% beyond 6$'$ for the EMOS.
Beyond 4$'$, the EPN tends to have 1--2\% larger values of the
contribution of the cooler component than the EMOS.
This difference reflects the discrepancy at 0.8--1 keV between the EPN and
the EMOS shown in Figure 2.
This energy band corresponds to an instrumental low-energy tail of the
Fe-L lines and the effect for the EPN is stronger than that of the EMOS.
The EPN has an additional uncertainty due to the position dependent energy resolution.
Considering that the EMOS spectra of $R>6'$ at 0.7--1.1 keV are well fitted by the
single temperature MEKAL model fit, 
 the larger amount of 1 keV component observed by EPN  may be caused by
the uncertainty of the response matrix of the EPN.
Thus, there is a systematic uncertainty of 1--2\% in the
fraction of the 1 keV component.
In summary, within $R<6'$, we detected at least two temperature
 components and outside $R>6'$, the upper limit of the second component
 is less than a few \% even using the EPN.

ASCA spectra of M~87 are also well fitted  with a sum of  two Raymond-Smith models (Raymond \& Smith 1977).
The temperature of the two components are 3.0 and 1.3 keV, and  the
ratio of the emission measure range from
0.2 to 2 within 10$'$ (Matsumoto et al. 1996), although the results of MEKAL model fit with ASCA spectra
are mostly consistent when considering the difference in spatial resolution (Finoguenov \& Jones 2000; Shibata 2000).
The contribution of the cooler component plotted in Figure 5 is 
much smaller than that derived from R-S fits.

\begin{figure}[]
\resizebox{\hsize}{!}{\includegraphics{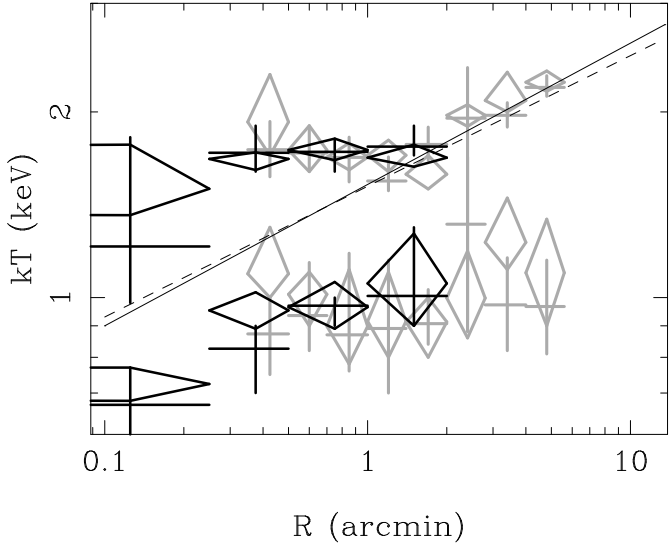}}
\caption{
The radial temperature profile obtained by fitting the spectra with
 a  two component MEKAL model. 
The meanings of symbols is the same as in Figure 3.
The solid (EMOS) and dashed line (EPN) corresponds to the best-fit power-law ration of 
the single component MEKAL fit, excluding the radio lobe regions.
}
\end{figure}

\begin{figure}[]
 \resizebox{\hsize}{!}{\includegraphics{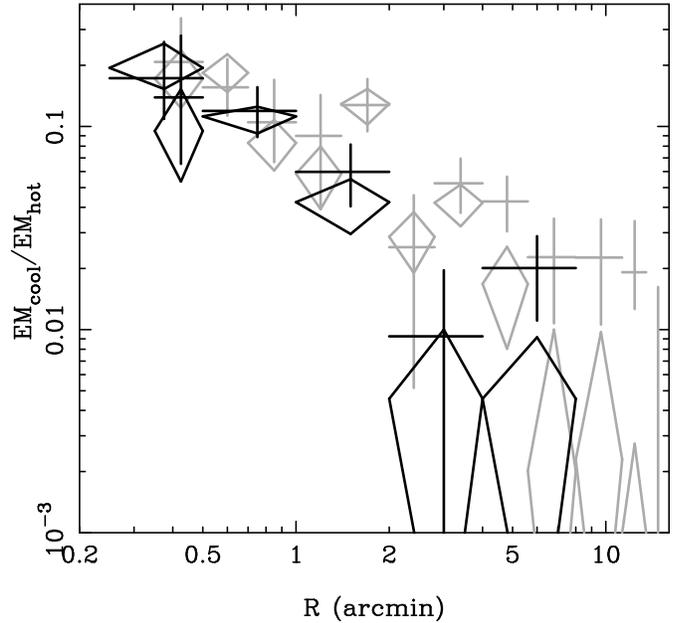}}
\caption{
The ratio of emission measure of the cool and hot temperature component,
assuming the temperature of the former to be 1.0 keV.
The meanings of symbols is the same as in Figure 3.
The discrepancies between the EPN and EMOS may be caused by instrumental
 problems in the EPN (see text).
}
\end{figure}

\subsection{Azimuthal temperature structure}

\begin{figure*}[]

\centerline{
\resizebox{5.2cm}{!}{\includegraphics{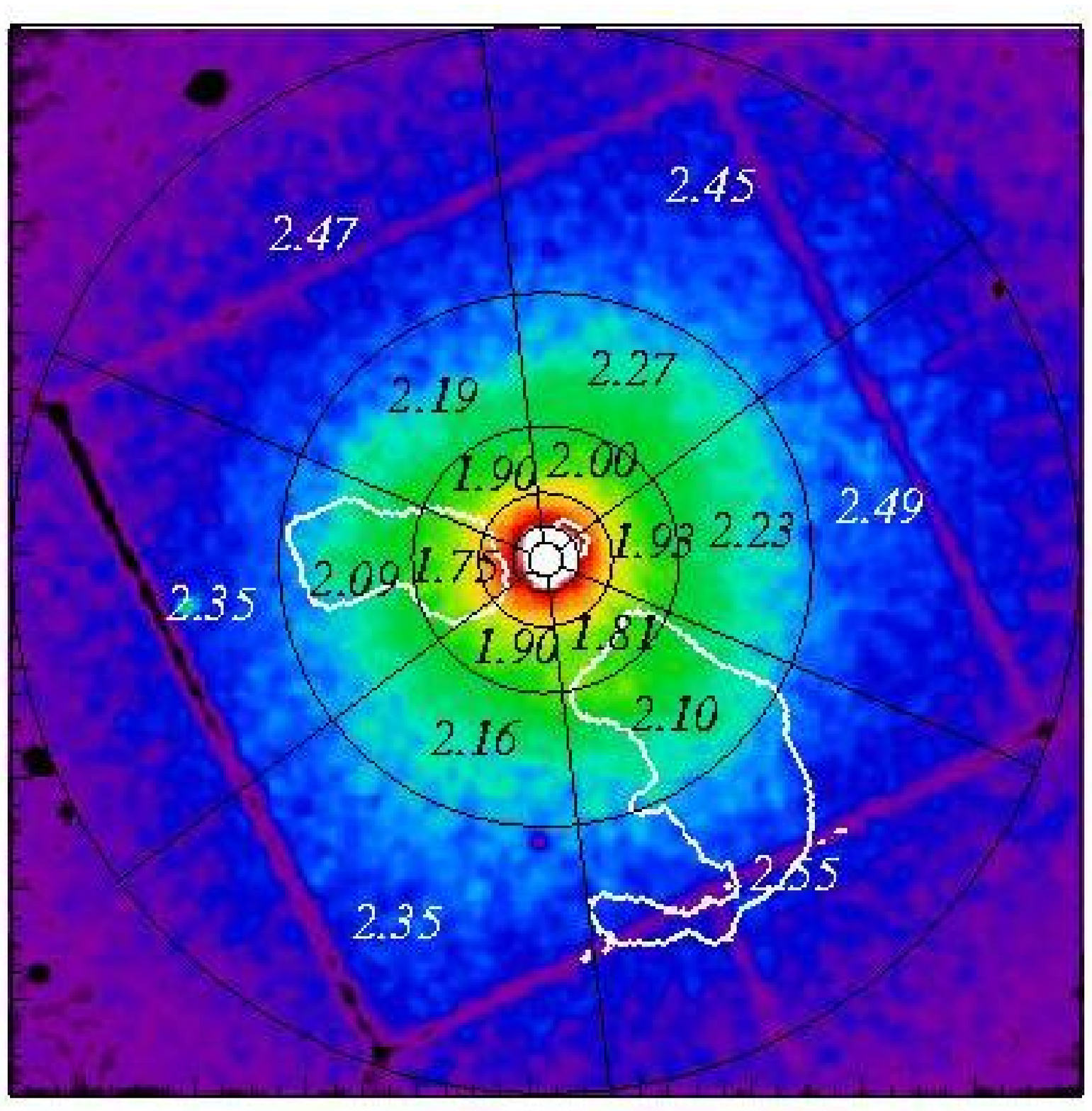}}
\hspace{0.3cm}
\resizebox{5.2cm}{!}{\includegraphics{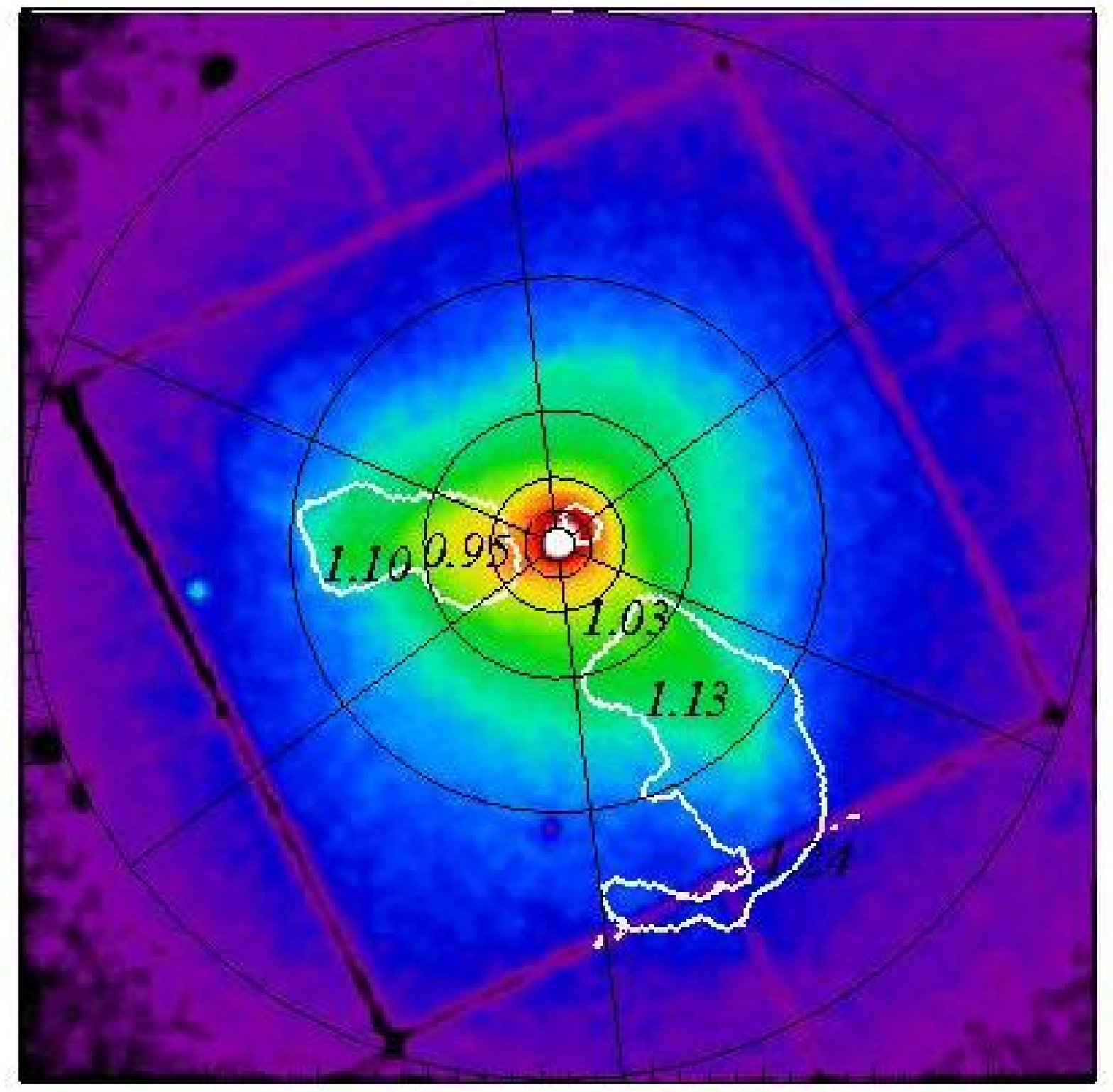}}
\hspace{0.3cm}
\resizebox{5.2cm}{!}{\includegraphics{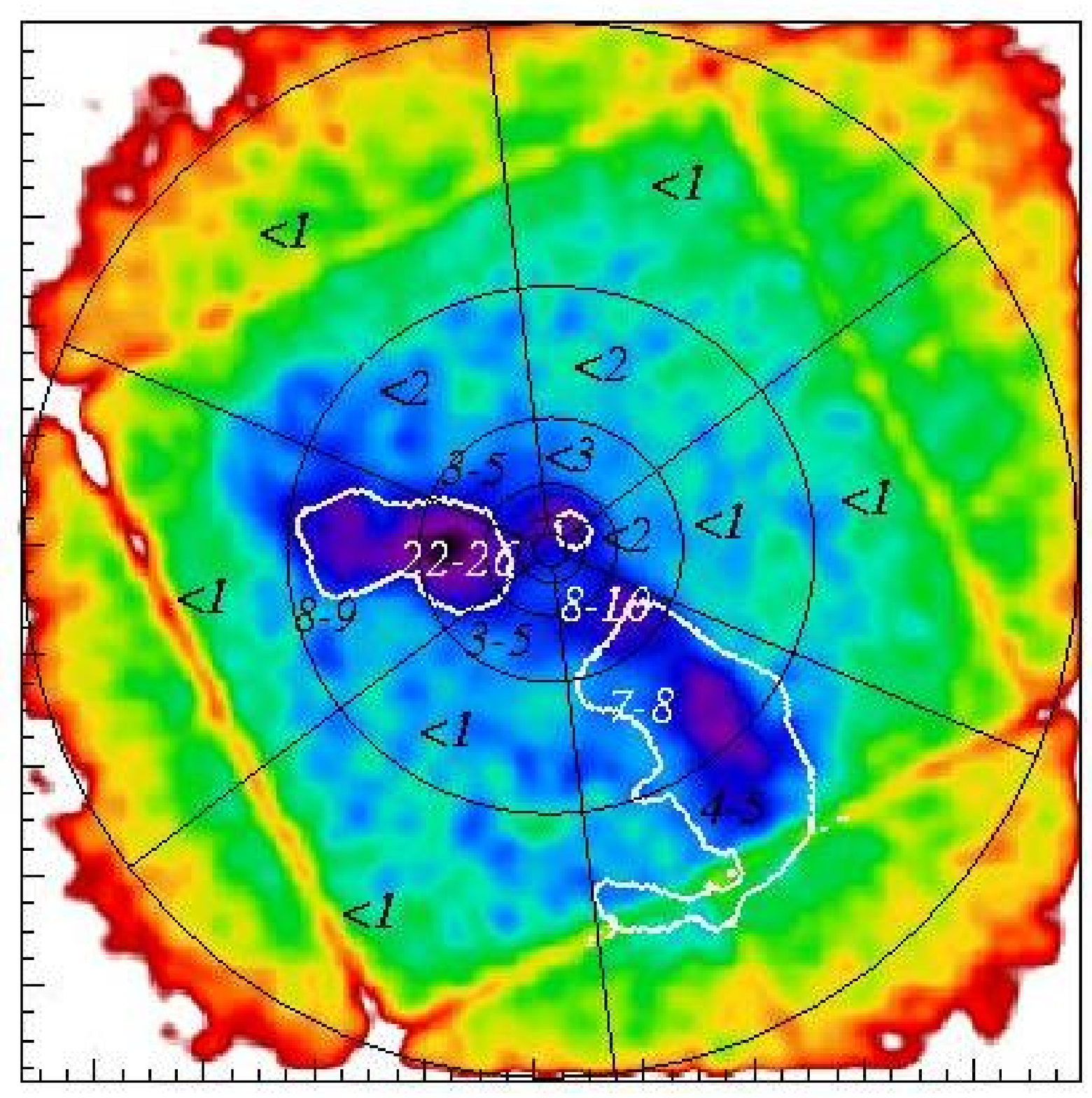}}
}
\centerline{
\resizebox{5.2cm}{!}{\includegraphics{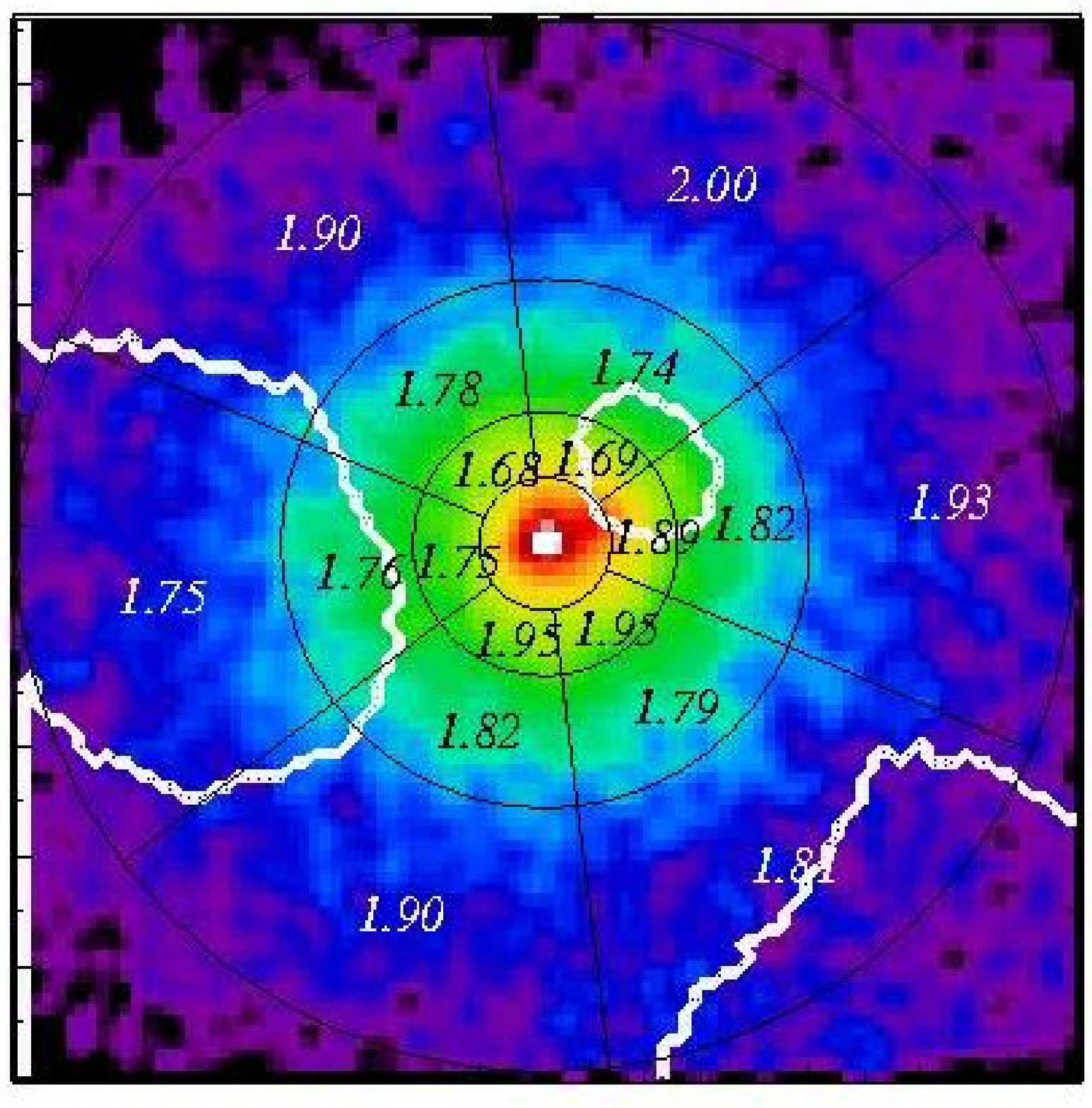}}
\hspace{0.3cm}
\resizebox{5.2cm}{!}{\includegraphics{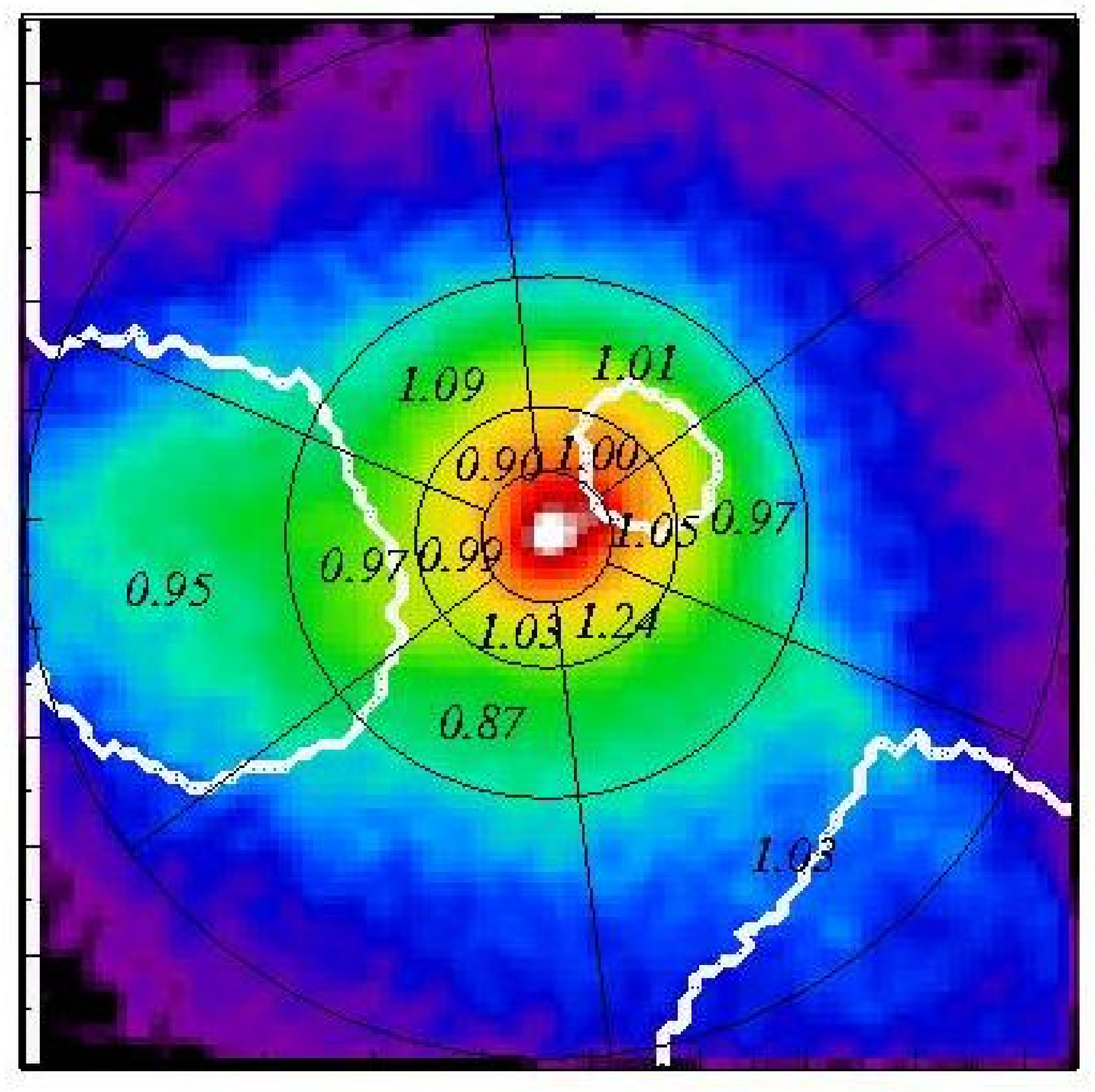}}
\hspace{0.3cm}
\resizebox{5.2cm}{!}{\includegraphics{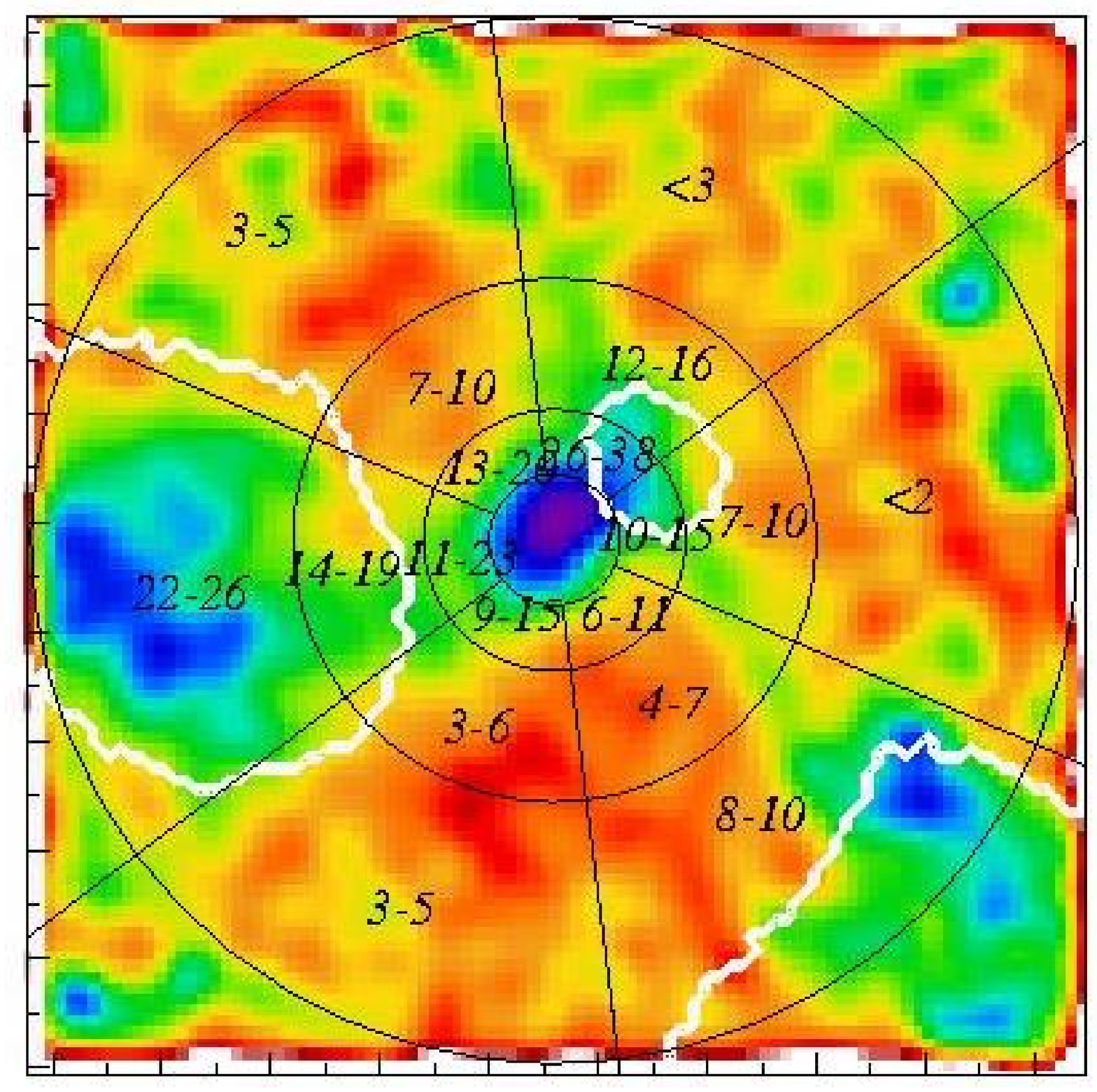}}
}

\caption{Hard (2--10 keV; left panels) and  Fe-L blend (0.6--1.2 keV;
 middle panels)  EMOS images, overlayed by the hard and soft
 components temperatures of the two MEKAL component model fit to  the projected
 sector regions, respectively.  The 90\% uncertainty of temperatures of
 the cool and hot components are typically 0.1 and 0.03 keV, respectively.
 The top panels and bottom panels are
 16$'\times$16$'$ and 4$'\times$4$'$ regions, respectively.
The right panels show the hardness ratio maps of 2--10keV/0.6--1.2keV (top)
and 1.0--1.2keV/0.8--1.0keV (bottom), overlayed  by the ratio of the
 emission measure of the soft to hard components in unit of percent.
White contours corresponds to the regions excluded from spectral
 analysis after \S3.3. }
\label{phispec}
\end{figure*}

In a second step,
in order to study the spatial distribution of the temperature
components in more detail, the EMOS spectra are accumulated within sector regions as
shown  in Figure \ref{phispec}.
The spectral fits used two temperature MEKAL model.
Although the spectra fitted in this section are not the deprojected ones, we
can constrain the spatial distribution of the temperature structure.
Figure \ref{phispec} shows 
EMOS images of the Fe-L (0.6-1.2 keV) and hard (2-10 keV) energy band
with the  temperatures obtained  for the two components.
It also shows  hardness ratio maps for the energy band ratios 0.6--1.2keV/2--10keV
and 1.0--1.2keV/0.8--1.0keV.
Although the latter is the most sensitive to the 1 keV components
(Figure 7), 
the statistics is not sufficient in the outer regions.
The ratios of the emission measure of the cool to hot components are
overlayed on the hardness ratio maps.

The hard band image is nearly spherically symmetric and 
no enhancements of the X-ray emission associated to the radio structure is seen.
The temperature of the hot components is also almost spherically symmetric
within 10\%.  It tends  to be slightly lower in the region around the radio lobes.

In contrast, the spatial distribution of the cool component is far from
symmetry, although  its temperature is determined to have an almost
constant value,  $\sim$ 1 keV.
Belsole et al. (2001) found that the softer regions agree well with the
radio structures.
Within $R<$1', the 1 keV component is detected from all the spectra,
but its contribution ranges from several \% to 30 \%.
In the radial zone $R$=1-2$'$, where 4 sectors are located outside
the radio lobes, we find a signature of the 1 keV component in  two
of these sectors but it is absent in the other two.
Outside 2$'$, we have only detected the 1 keV component
from the two sectors containing the radio lobes.

\subsection{Excluding the radio lobe regions}

For the further analysis,
we have filtered regions with excess emission related to the jet and radio lobes.
We used a spatial  filter, excluding those regions where the brightness  is larger by 15 \% than
the  azimuthally  averaged value (Figure 1, \ref{phispec}).
 The softer regions are not fully excluded within $2'$,
since it is difficult to filter whole the 1 keV component within 2$'$ 
due to its complicated structure and the limited spatial resolution of the XMM telescope.

Assuming spherical  symmetry,  deprojected spectra are calculated.
We also made deprojected spectra with larger shell width than those in Table 1.
We accumulated spectra within annular regions again and subtracted the contribution 
from the outer shells with thin shells width obtained before.
We then fitted the spectra with the single temperature and two temperature MEKAL model,
in the same way as in \S3.1.
In the following sections, we will only use the spectra filtered the
regions with enhance emissions.

\begin{table*}
\begin{tabular}[t]{rcccrrccrccrccr}
\hline
& & \multicolumn{3}{l}{1T MEKAL} &\multicolumn{5}{l}{2T MEKAL}  \\ 
\hline
   R     & &  $kT$       &$N_H$  &$\chi^2/\mu^a$ &   $kT1$       & $kT2$ & $N_H$  &$\chi^2/\mu$ & EM2/EM1$^b$  \\
(arcmin) & & (keV)  &\tiny{($10^{20}\rm{cm^{-2}}$)}      & &(keV)  & (keV) &\tiny{($10^{20}\rm{cm^{-2}}$)} \\
\hline
0.00-0.12&EPN & 0.69$^{+0.11}_{-0.07}$ & 2.0 (fix) & 49/56 & \\
0.12-0.25&EPN& 0.81$^{+0.15}_{-0.17}$ & 9.5$^{+7.5}_{-4.3}$ & 35/54 & \\
0.25-0.35&EPN& 1.19$^{+0.09}_{-0.19}$ & 3.0$^{+5.9}_{-2.7}$ & 72/54 & \\
0.35-0.50&EPN& 1.48$^{+0.11}_{-0.10}$ & 4.7$^{+3.0}_{-2.6}$ & 65/54 &1.77$^{+0.26}_{-0.22}$ & 0.89$^{+0.22}_{-0.17}$ &  2.8$^{+3.4}_{-2.8}$ & 55/52 &   0.139$^{+0.142}_{-0.072}$\\
0.50-0.70&EPN& 1.36$^{+0.05}_{-0.05}$ & 5.7$^{+2.7}_{-1.7}$ & 67/54 &1.71$^{+0.17}_{-0.13}$ & 0.84$^{+0.17}_{-0.10}$ &  2.7$^{+2.6}_{-2.6}$ & 46/52 &   0.180$^{+0.116}_{-0.068}$\\
0.70-1.00&EPN& 1.49$^{+0.03}_{-0.05}$ & 4.4$^{+1.7}_{-1.8}$ & 43/54 &1.72$^{+0.20}_{-0.17}$ & 1.05$^{+0.26}_{-0.22}$ &  2.3$^{+2.3}_{-2.3}$ & 35/52 &   0.085$^{+0.055}_{-0.043}$\\
1.00-1.40&EPN& 1.52$^{+0.07}_{-0.04}$ & 4.4$^{+1.6}_{-1.9}$ & 35/54 &1.64$^{+0.15}_{-0.11}$ & 0.91$^{+0.33}_{-0.26}$ &  3.3$^{+1.8}_{-2.0}$ & 30/52 &   0.059$^{+0.043}_{-0.039}$\\
1.40-2.00&EPN& 1.58$^{+0.06}_{-0.08}$ & 3.4$^{+1.7}_{-1.7}$ & 56/54 &1.82$^{+8.72}_{-0.18}$ & 1.12$^{+0.42}_{-0.27}$ &  1.7$^{+2.1}_{-1.7}$ & 46/52 &   0.060$^{+0.040}_{-0.030}$\\
2.00-2.80&EPN& 1.85$^{+0.09}_{-0.09}$ & 2.0$^{+1.2}_{-1.2}$ & 35/54 & &  && & 0.008$^{+0.020}_{-0.008}$\\
2.80-4.00&EPN& 2.03$^{+0.09}_{-0.09}$ & 2.2$^{+1.0}_{-0.9}$ & 70/54 &  &  &&  & 0.011$^{+0.016}_{-0.011}$\\
4.00-5.60&EPN& 2.14$^{+0.10}_{-0.09}$ & 1.9$^{+0.8}_{-1.1}$ & 74/54 & 2.20$^{+0.12}_{-0.11}$ & 1.0 (fix) &  1.5$^{+1.0}_{-1.0}$ & 65/53   & 0.018$^{+0.015}_{-0.015}$\\
\hline
0.0-0.25&EPN&0.87$^{+0.07}_{-0.05}$ &  5.8$^{+1.2}_{-2.8}$ & 53/54 &1.21$^{+0.61}_{-0.23}$ & 0.67$^{+0.10}_{-0.36}$ &  2.0$^{+10.2}_{-2.0}$ & 31/53 &   3.689$^{+11.461}_{-3.233}$\\
 0.25-0.50&EPN& 1.35$^{+0.06}_{-0.06}$ & 5.5$^{+1.9}_{-1.7}$ & 76/54 & 1.66$^{+0.15}_{-0.13}$ & 0.82$^{+0.10}_{-0.09}$ &  3.1$^{+2.0}_{-2.3}$ & 47/51 &   0.173$^{+0.088}_{-0.064}$\\
0.50-1.00&EPN& 1.47$^{+0.02}_{-0.04}$ & 4.5$^{+1.6}_{-1.0}$ & 84/54 &1.72$^{+0.11}_{-0.10}$ & 0.97$^{+0.13}_{-0.12}$ &  2.3$^{+1.4}_{-1.4}$ & 49/52 &   0.119$^{+0.040}_{-0.030}$\\
1.00-2.00&EPN& 1.55$^{+0.05}_{-0.04}$ & 4.1$^{+0.9}_{-1.1}$ & 71/54 &1.74$^{+0.16}_{-0.09}$ & 1.03$^{+0.26}_{-0.17}$ &  2.3$^{+1.2}_{-1.2}$ & 51/52 &   0.060$^{+0.022}_{-0.020}$\\
2.00-4.00&EPN& 1.97$^{+0.06}_{-0.05}$ & 2.2$^{+0.6}_{-0.6}$ & 91/54 &&&&&0.009$^{+0.010}_{-0.009}$\\
\hline
0.00-0.12&EMOS& 1.08$^{+0.26}_{-0.23}$ &  3.4$^{+2.2}_{-2.4}$ &  113.6/120\\
0.12-0.25&EMOS&  0.95$^{+0.05}_{-0.05}$ &  4.6$^{+1.0}_{-0.9}$ & 130.9/120\\
0.25-0.35& EMOS& 1.08$^{+0.03}_{-0.06}$ &  4.8$^{+2.1}_{-2.0}$ & 107.5/120\\
0.35-0.50&EMOS&  1.53$^{+0.09}_{-0.07}$ &  4.0$^{+1.6}_{-1.6}$ & 117.6/116 & 1.83$^{+0.33}_{-0.20}$ & 1.02$^{+0.28}_{-0.16}$ &  0.9$^{+1.6}_{-0.4}$ &  100.4/114  & 0.098$^{+0.079}_{-0.051}$ \\
0.50-0.70&EMOS&  1.38$^{+0.02}_{-0.04}$ &  4.4$^{+1.1}_{-0.7}$ & 149.0/116 & 1.83$^{+0.27}_{-0.13}$ & 1.08$^{+0.12}_{-0.08}$ &  2.1$^{+1.3}_{-1.3}$ &  111.1/114  & 0.179$^{+0.050}_{-0.040}$ \\
0.70-1.00&EMOS&  1.51$^{+0.04}_{-0.04}$ &  3.9$^{+0.7}_{-1.4}$ & 118.1/116 & 1.71$^{+0.09}_{-0.11}$ & 0.90$^{+0.20}_{-0.10}$ &  2.0$^{+1.1}_{-1.0}$ &  93.0/114  & 0.088$^{+0.028}_{-0.024}$ \\
1.00-1.40&EMOS&  1.56$^{+0.05}_{-0.06}$ &  3.4$^{+0.5}_{-1.2}$ &118.0/116 & 1.66$^{+0.14}_{-0.06}$ & 0.86$^{+0.34}_{-0.16}$ &  2.4$^{+1.0}_{-1.0}$ & 110.0/114  & 0.050$^{+0.023}_{-0.021}$ \\
1.40-2.00&EMOS&  1.59$^{+0.06}_{-0.04}$ &  2.9$^{+0.9}_{-0.9}$ &
 103.3/116& 1.68$^{+0.08}_{-0.06}$& 1.0 (fix) &  2.1$^{+1.0}_{-1.0}$& 99.5/114 & 0.035$^{+0.021}_{-0.020}$ \\
2.00-2.80&EMOS&  1.96$^{+0.07}_{-0.09}$ &  1.9$^{+0.6}_{-0.8}$ & 112.4/116 &  & & &  & 0.019$^{+0.011}_{-0.010}$ \\
2.80-4.00&EMOS&  2.10$^{+0.05}_{-0.06}$ &  1.2$^{+0.5}_{-0.4}$ & 121.3/116 &  & & & & 0.000$^{+0.007}_{-0.000}$ \\
4.00-5.60&EMOS&  2.19$^{+0.05}_{-0.05}$ &  1.1$^{+0.4}_{-0.4}$ &
 116.4/116 & &&& & 0.008$^{+0.008}_{-0.008}$ \\
\hline
0.00-0.25&EMOS&  0.95$^{+0.09}_{-0.09}$ &  4.0$^{+0.6}_{-0.7}$ & 176.9/120 & 1.50$^{+0.30}_{-0.10}$ & 0.72$^{+0.08}_{-0.02}$ &  2.2$^{+0.7}_{-1.0}$ &   104.2/114 & 0.488$^{+2.953}_{-0.276}$ \\
0.25-0.50&EMOS&  1.37$^{+0.04}_{-0.04}$ &  4.6$^{+1.0}_{-1.0}$ &  153.6/120&  1.68$^{+0.02}_{-0.08}$ & 0.95$^{+0.05}_{-0.05}$ &  2.7$^{+0.7}_{-0.6}$ &    113.7/114& 0.194$^{+0.062}_{-0.041}$ \\
0.50-1.00&EMOS&  1.46$^{+0.01}_{-0.01}$ &  4.2$^{+0.4}_{-0.5}$ &   196.4/116& 1.74$^{+0.06}_{-0.04}$ & 0.96$^{+0.14}_{-0.06}$ &  2.0$^{+0.7}_{-0.7}$ & 112.4/114   & 0.112$^{+0.013}_{-0.020}$ \\
1.00-2.00&EMOS&  1.58$^{+0.01}_{-0.02}$ &  3.1$^{+0.5}_{-0.4}$ & 150.2/116 & 1.69$^{+0.11}_{-0.09}$ & 1.05$^{+0.25}_{-0.15}$ &  2.2$^{+0.5}_{-0.6}$ & 135.4/114   & 0.042$^{+0.013}_{-0.013}$ \\
2.00-4.00&EMOS&  2.05$^{+0.03}_{-0.04}$ &  1.5$^{+0.3}_{-0.4}$ & 171.8/116 & & & & & 0.005$^{+0.005}_{-0.005}$ \\
4.00-8.00& EMOS&2.30$^{+0.04}_{-0.04}$ &  1.6$^{+0.2}_{-0.2}$ & 168.2/116 & & & & & 0.005$^{+0.005}_{-0.005}$ \\
\hline
\end{tabular}

a: degrees of freedom\\
b: Ratio of emission measure of the cooler and hotter component, when the temperature of the former is fixed to 1 keV.
\caption{Result of spectrum fitting of the deprojected spectra (regions
 with the enhanced X-ray emission associated to the radio structures are
 excluded)}
\end{table*}

\begin{figure}[]
\resizebox{\hsize}{!}{\includegraphics{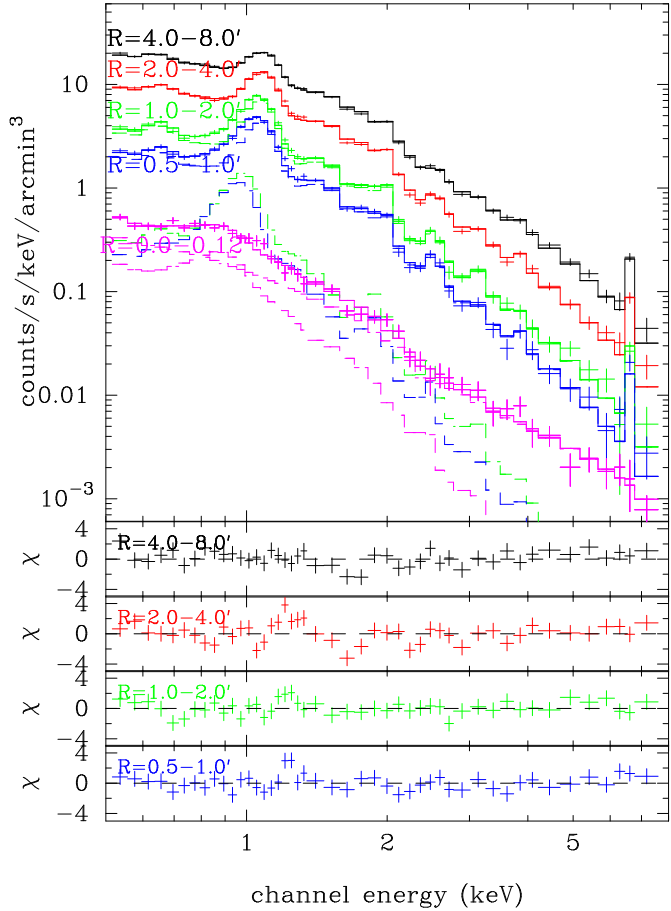}}
\caption{ Deprojected spectra of the EPN (lobes excluded). 
The spectrum at $R<0.12'$ is fitted with the double component model, consisting
a MEKAL model and a power-law component and those of R=0.5-1.0$'$ and 1.0-2.0$'$
are fitted with a two component MEKAL model.
Each component is plotted in dashed lines.
The other spectra are fitted with a MEKAL model.
The bottom panels show residuals of the fit. 
}
\label{pnspec}
\end{figure}

The results are summarized in Table 2.
After filtering the regions with excess emissions,
the emission measure of the 1 keV component is  reduced (Figure 5),
and the single temperature model gives reasonable fits at $R>2'$ (Table
2; Figure \ref{pnspec}).
 The remaining 1 keV components of the EPN beyond 2$'$ may be due to the
uncertainty of the instrumental low-energy tail of the Fe-L lines.
The radial temperature profile from the single temperature MEKAL model
fit  is
approximated as $kT=1.52^{+0.02}_{-0.02}R^{0.23^{+0.01}_{-0.01}}$ keV for the EMOS
and $kT=1.51^{+0.03}_{-0.03}R^{0.21^{+0.02}_{-0.01}}$ keV for the EPN (Figure 3),
where $R$ is in units of arcmin.

Within $R<2'$, the two temperature model still gives better fits than
the single temperature model, because of the complex structure of the 1 keV within the region.
When we adopted the hotter component temperature within $2'$, the radial
temperature profile is approximated as
$kT=1.69^{+0.09}_{-0.08}(1+(R/1.6'))^{0.115^{+0.011}_{-0.010}}$ keV and
$kT=1.68^{+0.04}_{-0.05}(1+(R/1.6'))^{0.100^{+0.017}_{-0.016}}$ keV 
for the EMOS and EPN, respectively.

There are some residuals at  $\sim$1.2 keV, which are common to all the deprojected shells.
Any multi-temperature model cannot explain the residual, because the continuum
and the Fe-K line are fitted well with a single MEKAL model.
When the Ni abundance is assumed to be a factor of 2
larger than Fe in  solar units, the residuals  become smaller
(Finoguenov et al. 2001; Matsushita et al. 2002). 
 This structure may also be due to a problem 
the Fe-L atomic data,  since the APEC code (Smith et al. 2001) do not
need such a high Ni abundance.

These results indicate that most of the space is occupied by a
component which is nearly spherically symmetric. 
Within relatively small regions associated to the radio structures,
there is an additional component whose temperature is about 1 keV.
In the following sections, we will study the temperature structure in
detail using the filtered spectra.

\section{The abundance profile of Fe}

\begin{figure}[]
\resizebox{\hsize}{!}{\includegraphics{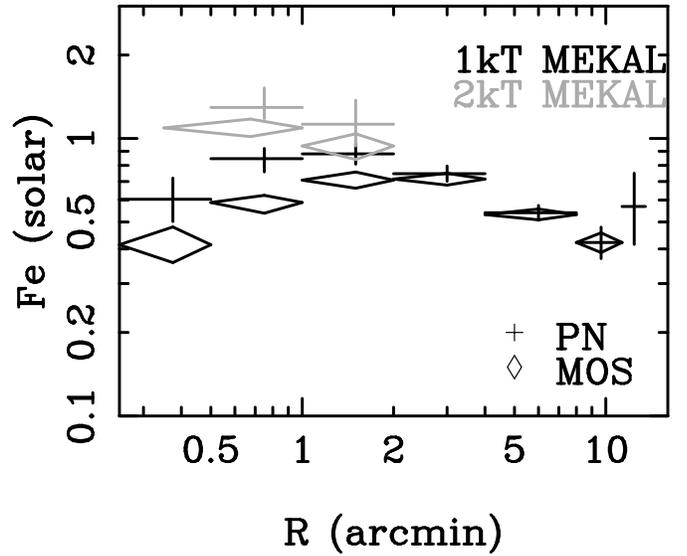}}
\caption{ Deprojected Fe profile of the EPN (crosses) and EMOS
 (diamonds) fitted with a single temperature MEKAL model (black) and the
 two temperature MEKAL model (gray).
}
\label{fe}
\end{figure}

In spectral fitting the derived temperature and element
abundances are partly dependents on each other.
This is particularly important for the Fe abundance, where the
spectral features of K- and L-shell lines with their temperature
dependent
relative importance  are fitted simultaneously.  A more detailed report
 on the abundances is given in Paper II
(Matsushita et al. 2002) and Finoguenov et al. (2001), however
we briefly discuss here the results on the Fe abundances in relation to
the derived temperature structure.
As derived from the projected spectra (B\"ohringer et al. 2001; Finoguenov
et al. 2001), the  Fe abundance shows a strong negative gradient (Figure \ref{fe}
).
The two temperature MEKAL fit gives significantly larger abundances than the single temperature MEKAL fit,
although the contributions of the cooler component are small.
For example,  at $R$=0.5-1', including of 10 \% of the cooler component in units of emission measure
changes Fe abundances by a factor 2. 
The reason for this is the change of the temperature of the hot
component, when the cold component is added, e.g. in the ring taken as
an example the temperature of the main component increases from 1.47 to
1.72 keV. For a fixed Fe line feature this leads to a higher abundance
required in the fit.
As a result, the  central abundance drop which is seen in the single MEKAL fit 
is not seen in the result of the two MEKAL fit.
This will be discussed further in paper II (Matsushita et al. 2002).

\section{Detailed temperature structure}

\subsection{Temperatures obtained from the Fe-L and the hard energy band}

There are some problems in the determination of the temperature through X-ray spectral fitting.
The first problem is  resonant line scattering, 
which should be important in the core of M~87 (Shigeyama 1998, B\"ohringer et al. 2001).
The second problem is the uncertainties in the Fe-L atomic data (e.g. Masai
1977; Matsushita et al. 2000).
The major uncertainty is the ionization and recombination rates of Fe,
which may change resulting temperature by 20--30\%.
(Masai 1997; Matsushita et al. 2000).
In order to avoid these problems, we exclude the Fe-L region below 1.6 keV and fit
the spectra again.  
In addition, to compare the temperature obtained to those from the Fe-L lines, we also fitted the data between
0.7 to 1.3  keV with a single temperature MEKAL or Raymond-Smith  model (hereafter R-S model).
 The R-S model is used for the comparison with the ASCA results.

The result is summarized in Table 3 and Figures  \ref{ktmosmekal}.
The derived temperatures using the hard energy band, $kT_{\rm hard}$, of the EPN and EMOS
are   consistent with  each other (Table 3).
At $R>2'$, they are consistent to those obtained using the whole energy band and
within $R$=0.35--2$'$ they  are consistent to the temperatures of the hotter component obtained in
\S3.1 and 3.3.
This indicates that within $R$=0.35--2$'$,  the 1.7 keV component really
exists and is the dominant phase in the ICM.

The results for the 
 Fe-L region using the MEKAL model, $kT_{\rm {MEKAL}}$, are consistent or slightly higher 
than $kT_{\rm {hard}}$ at $R>1'$ because of the residual structure around $\sim1.2$ keV.
Within $R<1'$, $kT_{\rm {MEKAL}}$ is lower than $kT_{\rm hard}$ 
 due to the contributions from the 1 keV components.

The  R-S model gives significantly lower temperatures than  $kT_{\rm {hard}}$ 
by $\sim$20\%.
As a result, when fitted with a R-S model, there remain residuals in the hard energy band.
To explain the residuals, we need an additional R-S component with a high temperature, 
$\sim$ 3 keV. { This is the reason of the comparable emission measure of
the two temperature R-S components obtained in the ASCA results
(Matsumoto et al. 1996)

\begin{table}[t]
 \begin{tabular}[t]{ccccc}
\hline
  &EPN & \multicolumn{3}{l}{EMOS}\\
   R     &   ${kT_{\rm hard}}^{a}$&  ${kT_{\rm hard}}^a$       &   ${kT_{\rm MEKAL}}^b$     &   ${kT_{\rm R-S}}^b$      \\
(arcmin) & (keV) & (keV) &  (keV) & (keV)  \\
\hline
0.00-0.12&                 &               & 1.08$^{+0.37}_{-0.27}$  &  1.01$^{+0.04}_{-0.02}$  \\
0.12-0.25&   &   & 1.03$^{+0.05}_{-0.05}$   & 1.01$^{+0.03}_{-0.03}$  \\
0.25-0.35  & 1.06$^{+0.34}_{-0.38}$ &   1.60$^{+0.39}_{-0.36}$   & 1.11$^{+0.06}_{-0.06}$   & 1.06$^{+0.03}_{-0.04}$ \\
0.35-0.50  & 1.78$^{+0.52}_{-0.38}$ &   1.64$^{+0.18}_{-0.17}$   & 1.59$^{+0.11}_{-0.09}$   & 1.38$^{+0.04}_{-0.05}$ \\
0.50-0.70  & 1.61$^{+0.16}_{-0.15}$ &   1.68$^{+0.11}_{-0.11}$   & 1.41$^{+0.03}_{-0.04}$   & 1.24$^{+0.03}_{-0.07}$ \\
0.70-1.00  & 1.59$^{+0.18}_{-0.16}$ &   1.60$^{+0.10}_{-0.10}$   & 1.55$^{+0.07}_{-0.05}$   & 1.33$^{+0.04}_{-0.03}$ \\
1.00-1.40  & 1.68$^{+0.18}_{-0.17}$ &   1.55$^{+0.10}_{-0.10}$   & 1.68$^{+0.08}_{-0.09}$   & 1.39$^{+0.02}_{-0.03}$ \\
1.40-2.00  & 1.74$^{+0.21}_{-0.15}$ &   1.58$^{+0.10}_{-0.10}$   & 1.73$^{+0.09}_{-0.10}$   & 1.39$^{+0.03}_{-0.03}$ \\
2.00-2.80  & 1.89$^{+0.15}_{-0.14}$ &   1.98$^{+0.09}_{-0.09}$   & 2.09$^{+0.16}_{-0.15}$   & 1.45$^{+0.12}_{-0.02}$ \\
2.80-4.00  & 2.12$^{+0.14}_{-0.13}$ &   2.05$^{+0.07}_{-0.07}$   & 2.10$^{+0.15}_{-0.12}$   & 1.53$^{+0.11}_{-0.09}$ \\
4.00-5.60  & 2.22$^{+0.21}_{-0.14}$ &   2.15$^{+0.08}_{-0.08}$   & 2.58$^{+0.32}_{-0.25}$   & 1.65$^{+0.11}_{-0.13}$ \\
5.60-8.00  & 2.41$^{+0.16}_{-0.15}$ &   2.29$^{+0.12}_{-0.07}$   & 2.46$^{+0.23}_{-0.19}$   & 1.71$^{+0.09}_{-0.11}$ \\
8.00-11.30 & 2.74$^{+0.21}_{-0.19}$ &  2.47$^{+0.09}_{-0.09}$  & 2.80$^{+0.34}_{-0.28}$   & 1.81$^{+0.29}_{-0.11}$ \\
11.3-13.50& 2.58$^{+0.36}_{-0.30}$ & 2.61$^{+0.06}_{-0.05}$ &2.65$^{+0.23}_{-0.19}$    & 2.04$^{+0.13}_{-0.14}$ \\
13.5-16.00& 2.55$^{+0.20}_{-0.18}$\\
\hline
 \end{tabular}
\\
a:  $kT$ obtained by fitting the spectra above 1.6 keV.\\
b:  $kT$ obtained by fitting the spectra between 0.7 to 1.3  keV.\\
\caption{Result of spectrum fitting of the deprojected spectra (radio lobes excluded)}
\end{table}
\begin{figure}[]
\resizebox{\hsize}{!}{\includegraphics{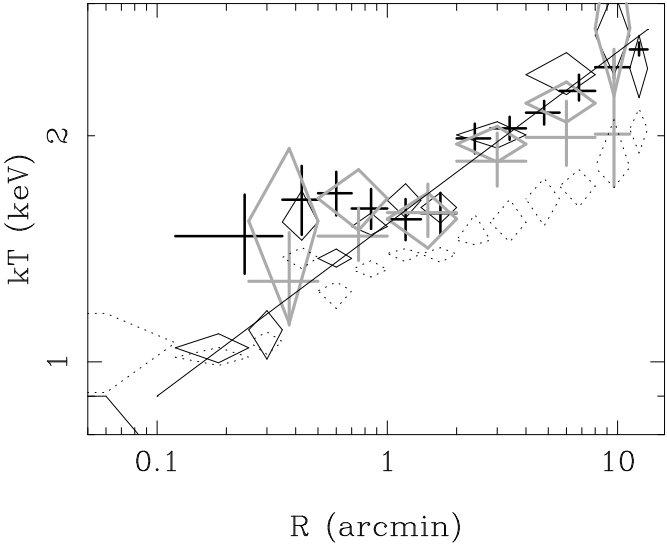}}
\caption{Deprojected radial profile of the temperature obtained with  EMOS, 
by fitting the spectra of 0.7--1.3 keV (black  diamonds), above 1.6 keV 
(black crosses), 1.8-2.1 keV (gray crosses) and 2.3-2.7 keV (gray
 diamonds) with a MEKAL model. 
 The results obtained by R-S model using 0.7--1.3 keV are also shown (dotted diamonds).
The solid line corresponds to the best fit regression line for the MEKAL
model using the whole energy band of the EMOS.
}
\label{ktmosmekal}
\end{figure}

\subsection{Temperature obtained from line ratios}

The ratio of He-like $K_\alpha$ to H-like $K_\alpha$ line strengths is  sensitive to the gas temperature.
Since the energy resolution of the EMOS is better than that of the EPN, we used only  EMOS data here.
We fitted the spectra within the energy range
of 1.8-2.1 keV, and 2.3-2.7 keV with a MEKAL model.
We denote the temperature obtained from the former energy range as $kT_{\rm Si}$ and 
from the latter as $kT_{\rm S}$.
The contribution of Fe-L lines at the energy of the He-like $K_\alpha$ line  of Si  is considerable,
and the He-like line contains several strong components. 
The result is summarized in Table 4.

Figure \ref{ktmosmekal}  also compares the temperatures,  $kT_{\rm Si}$ and $kT_{\rm S}$.
These temperatures  agree well within 10\%  with 
 the deprojected temperature of the corresponding shell, although
$kT_{\rm Si}$ may  have some uncertainty, since
around the Si He-K lines,  Fe-L and Ni-L lines are also important.
Furthermore, there is an instrumental Si edge in the response matrix and
a strong instrumental Si line, which may cause the discrepancy between
the EPN and EMOS at 2 keV shown in Figure 2.


We also compare the model and data spectra of the Si and S region in Figure \ref{sisspec}.
The two temperature R-S model with the same abundance ratio,
cannot fit the Si and S line ratios, while
a single temperature MEKAL model fits the lines well.
In order to explain the observed  line ratios with the two component R-S model,
the harder spectral component should have an abundance of
Si and S a factor of $3\sim5$ larger than the softer component.

\begin{table}
  \begin{tabular}[t]{ccccccccc}
\hline
          &\multicolumn{2}{l}{1.8-2.1keV} & \multicolumn{2}{l}{2.3-2.8keV}\\
 $R$     & $kT_{\rm{Si}}$ & $\chi^2/\mu^a$ & $kT_{\rm{S}}$ & $\chi^2/\mu^a$\\
(arcmin) & (keV)  &  & (keV) &\\
\hline
0.25-0.50& 1.28$^{+0.21}_{-0.16}$ &   19.2/19 &  1.54$^{+0.38}_{-0.42}$ & 18.4/31\\
0.50-1.00& 1.47$^{+0.13}_{-0.11}$ &   9.3/19 &  1.65$^{+0.15}_{-0.15}$ & 14.6/31\\
1.00-2.00& 1.58$^{+0.15}_{-0.11}$ &   17.5/19 &  1.55$^{+0.13}_{-0.13}$ & 31.2/31\\
2.00-4.00& 1.85$^{+0.17}_{-0.14}$ &   12.8/19 &  1.95$^{+0.11}_{-0.10}$ & 20.4/31\\
4.00-8.00& 1.99$^{+0.23}_{-0.16}$ &   9.0/19 &  2.21$^{+0.15}_{-0.12}$ & 23.0/31\\
8.00-11.30& 2.01$^{+0.60}_{-0.30}$ &   12.7/19 &  2.78$^{+1.26}_{-0.52}$ & 19.9/31\\
\hline
 \end{tabular}
\\
a: degrees of freedom
\caption{Result of spectral fit of EMOS data, using the energy band around the Si and S lines with a MEKAL model}
\end{table}

\begin{figure}[]
\resizebox{\hsize}{!}{\includegraphics{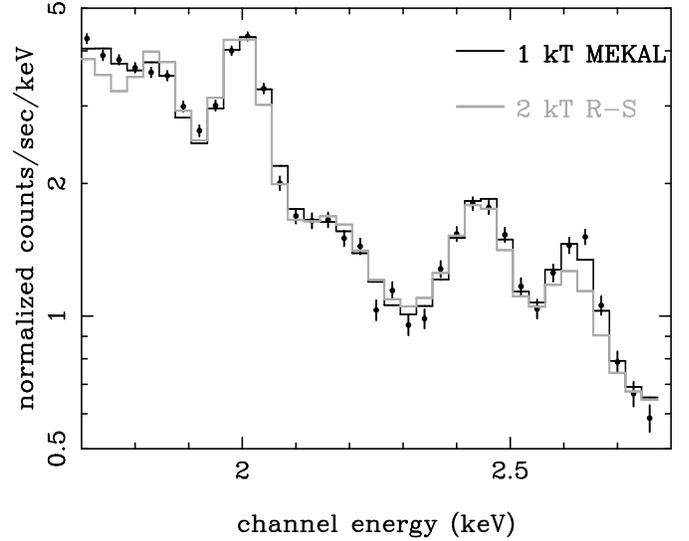}}
\caption{
Deprojected EMOS spectrum within R=1-2.8$'$ in the energy range of Si and S lines,
fitted with a MEKAL model  (black  line) or
 1.4 keV + 3.3 keV R-S model (gray line).
}
\label{sisspec}
\end{figure}

Since the deprojected data have larger uncertainties,
 we have also used the S line ratios of the annular spectra.
The strength of the  K$\alpha$ lines are obtained by fitting the annular
 spectra within 2.2--2.8 keV with a bremsstrahlung and two gaussians.
We calculated the projected S line ratio using the deprojected
 temperature and density (\S7) profiles and compared with the observed
 ones.
We used  two temperature relations derived in \S3.3.
One is the best fit relation of the single MEKAL model fit
and the other relation is derived when we adopted the hotter component
 temperature of the two temperature MEKAL fit within $R<2'$.
Outside 1$'$, the observed ratios agree well with the calculated
profiles from single temperature model fit (Figure \ref{sk_mekal}).
Comparing these results with a profile calculated using a temperature profile shifted by
10\% to the data, we can conclude that the temperature profile
inferred from the observed S line ratios is consistent
within several \% with the deprojected temperature profile.
Within 1$'$, the observed ratios are close to with those calculated using
the hotter component temperature of the two temperature MEKAL fit, but smaller than those from the single
temperature fit.
This result indicates that S lines are dominated by those of the hotter component
 within 2$'$.

\begin{figure}[]
\resizebox{\hsize}{!}{\includegraphics{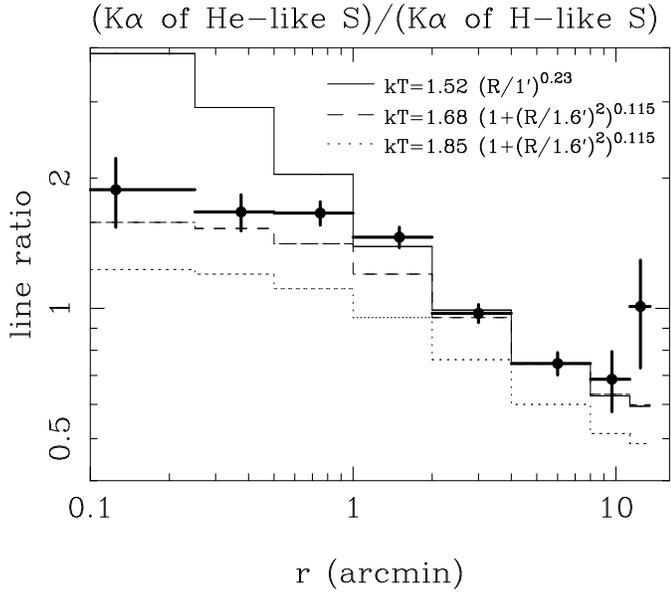}}
\caption{Projected line ratios (EMOS) between K$\alpha$ of  hydrogen and helium
 like S (closed circles). Errors correspond to 68\% confidence level.
The solid and dashed lines represent the results using the observed
 temperature profile from the single temperature MEKAL model
and that of the hotter component from the two temperature MEKAL model, respectively.
The assumed radial temperature profiles are written in the figure.
    We also plotted the profile for temperatures
 shifted by 10\% (dotted line)
}
\label{sk_mekal}
\end{figure}

We then compared the strength of the  Fe-L and Fe-K lines as follows.
As for the S line ratios,  the line brightness profile of Fe-K is
calculated using the deprojected results.
Within 2$'$, we adopted the Fe abundance obtained from the two
temperature model fit.
The Fe abundance from the deprojected data is mainly determined from 
 the Fe-L lines.  For a given temperature,  this analysis gives a  comparison between the Fe-L and
 Fe-K spectral signatures.
Outside 0.5$'$, the observed profile of Fe-K is only 20\% larger than the calculated profile
(Figure \ref{FeK}).
Since the Fe-L atomic data may have uncertainties of 20-30\% (e.g. Masai
et al. 1998; Matsushita et al. 2001),
the difference may be due to the uncertainties in the Fe-L atomic data.
If the difference is caused by the temperature structure, it is explained by
only 10\% larger temperatures.

\begin{figure}[]
\resizebox{\hsize}{!}{\includegraphics{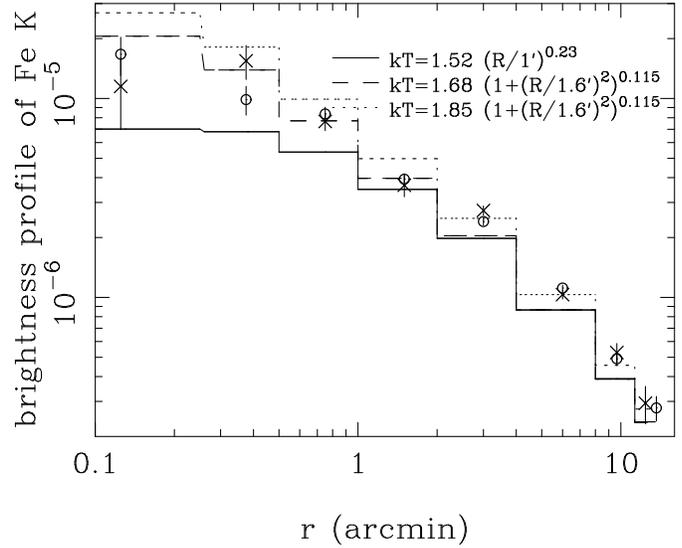}}
\caption{
The projected radial profile of the line brightness of
 Fe-K$\alpha$ of the EMOS (open circles) and the EPN (crosses).
The solid and dashed lines represent the results using the
 temperature profile from the single temperature MEKAL model
and that of the hotter component from the two temperature MEKAL model,
 respectively.
 The dotted lines corresponds to 10\% higher temperatures.
Errors represent the 68 \% confidence level. 
}
\label{FeK}
\end{figure}

In summary, excluding the regions associated to the radio structures,
at R$>$1$'$, $kT_{\rm{hard}}$, $kT_{\rm{Si}}$ and $kT_{\rm{S}}$ of a
given deprojected spectrum agree within 10\%.
$kT_{\rm{MEKAL}}$ also agrees well with $kT_{\rm{hard}}$, although
$kT_{\rm{Fe-L}}$ depends on the plasma emission code (e.g Masai 1997).
These results  indicate that at least at $R>1'$, the ICM at a given radius is
dominated by  a single temperature component whose temperature is 1.7
keV at $R$=1$'$ and 2.5 keV at $R$=15$'$.
The consistency between Fe-L and Fe-K also supports the single phase
temperature structure.
The difference of these temperatures within 1$'$ reflects that the
contribution of the 1 keV component become significant.

Moreover, the temperature obtained from the MEKAL model seems to be reasonable at least in
the obtained temperature range, when ignoring the residual structure at 1.2 keV.


\subsection{Spectral fits with the APEC plasma code}

A new plasma code, APEC (Smith et al. 2001), is now available in
XSPEC.
The residual structures around the Fe-L emission that occurred in the
fits with the MEKAL code 
disappeared with the use of the APEC code for Centaurus clusters (Sanders \& Fabian 2001).

In this paper, we used the version 1.0 of the APEC in the version 11.0.1
of the XSPEC.
In figure \ref{APEC_model}, we show ratios of a single temperature plasma with 
solar abundance calculated with the APEC and MEKAL models, when observed
with the EMOS.
Within the energy resolution of the CCD detectors, the dominant change is not
in the Fe-L lines, but in the  K$\alpha$ lines.
Using the APEC code, the strength of the  K$\alpha$ lines of
hydrogen like ions, which are the most fundamental lines,  
decrease by several tens of \%, and the temperature dependence
also changes.
The 6.7 keV Fe-K lines  decrease,  by a factor of 2 at 1 keV and 30\%
at 4 keV.
The changes of the K$\alpha$ lines of helium like ions of Si, S, Ar and Ca are smaller.
 
\begin{figure}[]
\resizebox{\hsize}{!}{\includegraphics{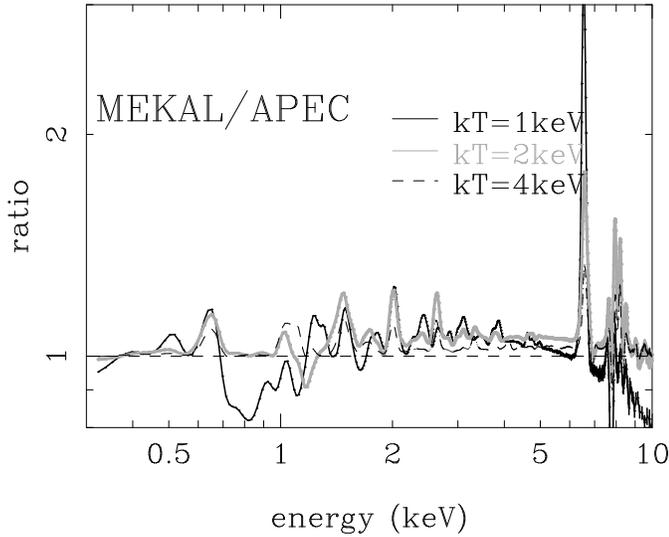}}
\caption{
Ratio of  theoretical spectra of MEKAL to APEC code for a plasma 
 temperature of 1 keV (black solid line), 2 keV (gray solid line) and 4
 keV (black dashed line) when observed with the EMOS. Abundances are assumed to have solar values.
}
\label{APEC_model}
\end{figure}

Figure \ref{mos_apec} shows a representative EMOS spectrum of M 87
fitted with the  MEKAL and APEC model.
As already shown, the single temperature MEKAL model can well fit the spectra including
the strength of Fe-L, Fe-K and S lines.
The APEC model gives a better fit  for the Fe-L/Mg-K structure around 1.2--1.5 keV and
the Fe-L/Si-K structure around 1.8 keV.
However, the single temperature APEC model cannot fit the Fe-K and Fe-K, 
as well as the ratio of K$\alpha$ lines of S simultaneously, although the fit to the Fe-L
complex and the continuum is good.

\begin{figure}[]
\resizebox{\hsize}{!}{\includegraphics{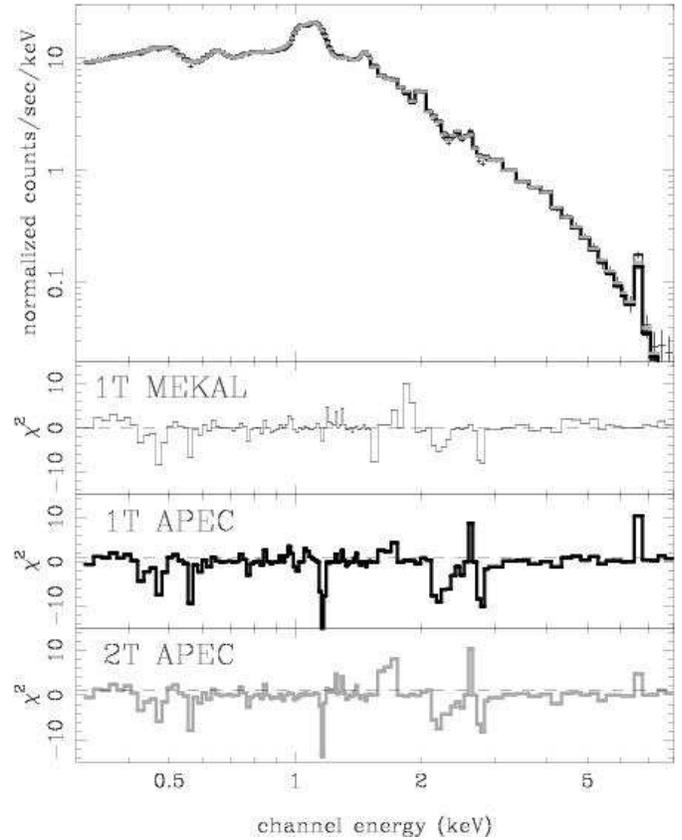}}
\caption{
The deprojected spectrum of EMOS within $R$=4-8$'$ (crosses) fitted with a
 MEKAL model (thin black lines), an APEC model (bold black lines) and a two
 temperature APEC model (bold gray lines). The bottom panels shows the
 contributions to $\chi^2$.
}
\label{mos_apec}
\end{figure}
   
\begin{table*}[]
 \begin{tabular}[t]{cccccccc}
 &\multicolumn{3}{c}{0.3--10keV} & 0.7--1.3keV & 1.6--10keV & 1.8-2.1keV
  & 2.3-2.8keV\\
 $R$     & kT    & $N_H$  &$\chi^2/\mu^a$ &  $kT$& $kT$ &$kT_{\rm{Si}}$ &
  $kT_{\rm{S}}$ \\
(arcmin) & (keV) & $10^{20}\rm{cm^{-2}}$ & & (keV)& (keV)& (keV)& (keV)\\
0.00-0.12& 1.04$^{+0.20}_{-0.15}$ &  1.6$^{+1.1}_{-1.0}$ & 111.9/116& 1.05$^{+0.25}_{-0.35}$& 1.36$^{+0.98}_{-0.30}$ \\
0.12-0.25& 1.02$^{+0.01}_{-0.08}$ &  4.0$^{+0.6}_{-0.7}$ & 140.6/116& 1.05$^{+0.02}_{-0.06}$& 1.29$^{+0.29}_{-0.31}$ \\
0.25-0.35& 1.02$^{+0.08}_{-0.05}$ &  6.6$^{+1.9}_{-2.2}$ & 106.1/116& 1.06$^{+0.11}_{-0.08}$& 1.53$^{+0.42}_{-0.50}$ \\
0.35-0.50& 1.61$^{+0.03}_{-0.10}$ &  3.6$^{+1.9}_{-1.9}$ & 109.0/116& 1.59$^{+0.09}_{-0.09}$& 1.71$^{+0.24}_{-0.18}$ \\
0.50-0.70& 1.34$^{+0.01}_{-0.02}$ &  5.0$^{+1.8}_{-0.8}$ & 148.3/116& 1.35$^{+0.01}_{-0.02}$& 1.80$^{+0.13}_{-0.11}$ \\
0.70-1.00& 1.54$^{+0.03}_{-0.02}$ &  3.4$^{+1.3}_{-0.6}$ & 120.3/116& 1.58$^{+0.04}_{-0.04}$& 1.73$^{+0.13}_{-0.09}$ \\
1.00-1.40& 1.62$^{+0.02}_{-0.05}$ &  3.3$^{+1.0}_{-0.8}$ & 111.0/116& 1.63$^{+0.04}_{-0.04}$& 1.71$^{+0.10}_{-0.09}$ \\
1.40-2.00& 1.63$^{+0.01}_{-0.03}$ &  1.9$^{+1.0}_{-0.8}$ & 92.8/116& 1.64$^{+0.02}_{-0.04}$& 1.72$^{+0.07}_{-0.09}$ \\
2.00-2.80& 1.92$^{+0.03}_{-0.02}$ &  1.8$^{+0.4}_{-0.4}$ &124.1/116 & 1.88$^{+0.10}_{-0.08}$& 2.06$^{+0.08}_{-0.07}$ \\
2.80-4.00& 2.12$^{+0.01}_{-0.03}$ &  1.1$^{+0.3}_{-0.3}$ &157.1/116 & 1.92$^{+0.07}_{-0.09}$& 2.13$^{+0.06}_{-0.06}$ \\
4.00-5.60& 2.13$^{+0.04}_{-0.03}$ &  1.0$^{+0.3}_{-0.3}$ & 123.3/116& 2.09$^{+0.05}_{-0.07}$& 2.20$^{+0.07}_{-0.06}$ \\
5.60-8.00& 2.32$^{+0.02}_{-0.05}$ &  1.8$^{+0.2}_{-0.4}$ &162.5/116 & 2.10$^{+0.05}_{-0.08}$& 2.35$^{+0.09}_{-0.07}$ \\
8.00-11.3& 2.54$^{+0.07}_{-0.06}$ &  1.7$^{+0.4}_{-0.3}$ & 169.4/116& 2.44$^{+0.17}_{-0.26}$& 2.48$^{+0.09}_{-0.08}$ \\
11.3-13.5& 2.67$^{+0.04}_{-0.03}$ &  1.3$^{+0.1}_{-0.3}$ &308.6/116 & 2.60$^{+0.10}_{-0.17}$& 2.59$^{+0.03}_{-0.05}$ \\
0.00-0.25& 1.04$^{+0.01}_{-0.06}$ &  3.2$^{+0.6}_{-0.4}$ & 180.9/116&
  1.02$^{+0.04}_{-0.04}$& 1.52$^{+0.21}_{-0.27}$ & 1.38$^{+0.51}_{-0.44}$ & 1.59$^{+3.89}_{-0.68}$ \\
0.25-0.50& 1.44$^{+0.04}_{-0.06}$ &  4.6$^{+1.0}_{-0.9}$ & 148.0/116& 1.34$^{+0.02}_{-0.03}$& 1.71$^{+0.22}_{-0.11}$& 1.39$^{+0.33}_{-0.12}$  & 1.64$^{+0.52}_{-0.45}$\\
0.50-1.00& 1.51$^{+0.02}_{-0.02}$ &  4.4$^{+0.4}_{-0.5}$ & 201.5/116& 1.50$^{+0.03}_{-0.03}$& 1.76$^{+0.08}_{-0.05}$ & 1.71$^{+0.29}_{-0.13}$  & 1.87$^{+0.19}_{-0.22}$\\
1.00-2.00& 1.62$^{+0.01}_{-0.02}$ &  2.7$^{+0.4}_{-0.5}$ & 126.0/116& 1.63$^{+0.02}_{-0.02}$& 1.73$^{+0.03}_{-0.05}$& 1.95$^{+0.20}_{-0.24}$  & 1.71$^{+0.17}_{-0.15}$  \\
2.00-4.00& 2.04$^{+0.02}_{-0.01}$ &  1.3$^{+0.2}_{-0.2}$ & 213.4/116&
  1.97$^{+0.04}_{-0.06}$& 2.11$^{+0.04}_{-0.04}$ & 2.20$^{+0.43}_{-0.13}$  & 2.19$^{+0.18}_{-0.10}$\\
4.00-8.00& 2.24$^{+0.02}_{-0.02}$ &  1.6$^{+0.1}_{-0.2}$ & 219.5/116&
  2.10$^{+0.04}_{-0.04}$& 2.31$^{+0.04}_{-0.04}$ & 2.29$^{+0.41}_{-0.23}$  & 2.56$^{+0.14}_{-0.15}$ \\
 \end{tabular}
\\a: degrees of freedom
\caption{Result of EMOS spectrum fitting of the deprojected spectra using
 a APEC model (lobe excluded)}
\end{table*}

Using the APEC code, we fitted the deprojected spectra of the EMOS
 in the same way as described in \S3.3, and \S5.1-2.
Figure \ref{ktvapec} and  Tables 5 summarize the temperatures derived
 with the APEC code.
 Due to the inconsistency of  the S line ratios and  the strength of the Fe-K line,
the reduced $\chi^2$ is systematically worse at $R>2'$ compared to those
 obtained from the MEKAL  model fit.
The temperatures obtained from the whole energy band, the Fe-L complex and
 the hard energy band using the APEC code 
 are consistent with those obtained using the MEKAL model.
The derived Fe abundances, which are mainly determined by Fe-L lines, are also consistent with
those from MEKAL model (figure \ref{feapec}).
In contrast,  the temperatures from Si and S energy band are 10\% larger, due
 to  the changes of the strength of the K$\alpha$ lines of hydrogen like ions.

The projected radial profile of the S line ratios are also compared with the 
calculated profile from the deprojected data (Figure \ref{sk_apec} ).
Although the deprojected temperatures cannot reproduce the observed line
ratios, the difference is only 10\% in temperature.
Therefore,  as in the case with MEKAL model,
temperatures derived from whole energy band, the Fe-L, the hard band
Si, and S energy band agree within 10\%.

\begin{figure}[]
\resizebox{\hsize}{!}{\includegraphics{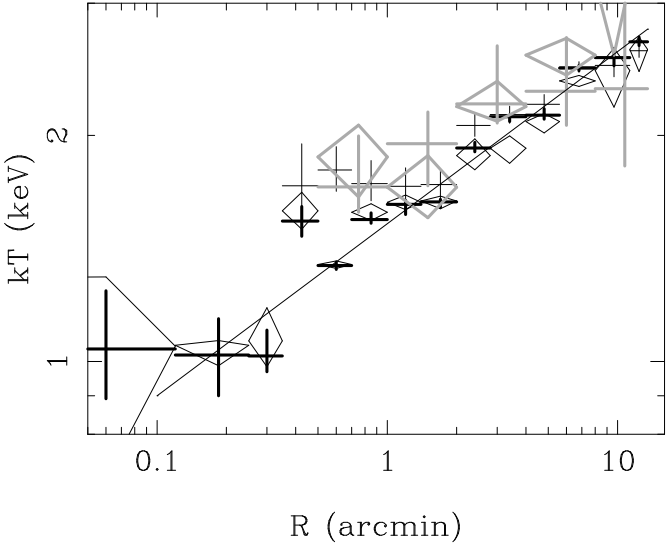}}
\caption{
Deprojected radial profile of the temperature obtained with  EMOS, 
by fitting the spectra of whole energy band (bold black crosses),
 0.7--1.3 keV (black diamonds), above 1.6 keV 
(black crosses), 1.8--2.1 keV (gray crosses) and 2.3--2.7 keV (gray
 diamonds) with a APEC model. 
The solid line corresponds to the best fit regression line for the MEKAL
model using the whole energy band of the EMOS.
}
\label{ktvapec}
\end{figure}

\begin{figure}[]
\resizebox{\hsize}{!}{\includegraphics{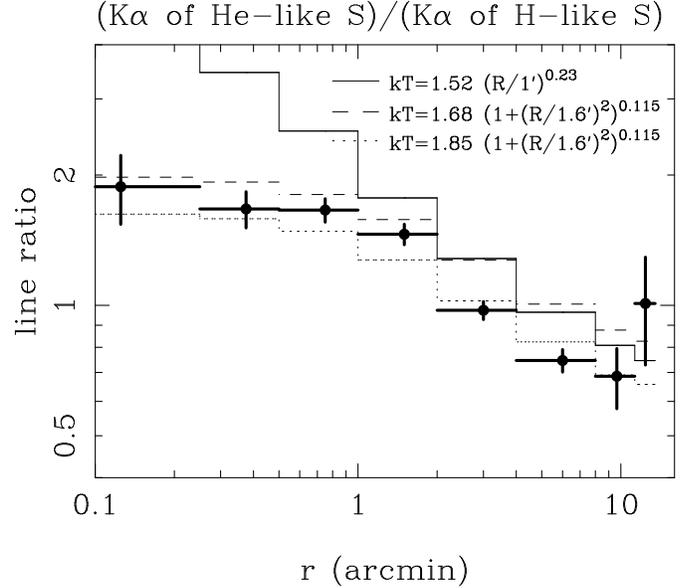}}
\caption{Projected line ratios (EMOS) between K$\alpha$ of  hydrogen and helium
 like S (closed circles). Errors correspond to 68\% confidence level.
Solid line and dashed lines represent the results 
  from  deprojected temperature profiles using APEC model,
 $kT_{\rm{whole}}$ and $kT_{\rm{hard}}$, respectively.
    We also plotted the profile for temperatures
 shifted by 10\% (dotted line)
}
\label{sk_apec}
\end{figure}

\begin{figure}[]
\resizebox{\hsize}{!}{\includegraphics{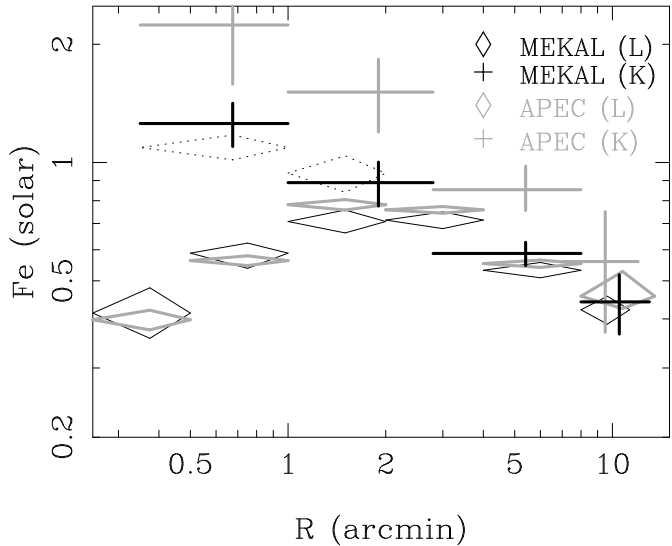}}
\caption{
Deprojected profile of Fe abundances obtained from the whole energy band
 (diamonds) and above 2 keV (crosses), using the single temperature
 MEKAL model (black),  APEC model (gray) and the two temperature
 MEKAL model (dotted).
}
\label{feapec}
\end{figure}

However, there is a discrepancy between the Fe-L and Fe-K strength.
Figure  17 also shows the Fe abundances obtained from the deprojected spectra above 2
keV, fitted with a single temperature MEKAL and APEC model.
We denote the derived Fe abundances from $>2$ keV, as Fe-K$_{\rm MEKAL}$ and Fe-K$_{\rm
APEC}$, and those from the whole energy band as $Fe-L_{\rm MEKAL}$ and
Fe-L$_{\rm APEC}$ for the MEKAL and APEC codes, respectively.
Outside $R>1'$, Fe-K$_{\rm MEKAL}$, Fe-L$_{\rm MEKAL}$, and Fe-L$_{\rm APEC}$ agree
 well  each other. Within $R<1'$,  Fe-K$_{\rm MEKAL}$  are consistent
 with Fe abundances obtained from the two temperature MEKAL model.
However,  Fe-K$_{\rm MEKAL}$ are systematically larger by $\sim$60\%.
The strength of the projected Fe-K lines are almost a factor of 2 larger than the
values calculated   from the deprojected results (Figure \ref{fekapec}). 
In order to obtain the observed Fe-K line profile, the temperature should be
shifted  by 30\%  compared to the deprojected temperatures.

Adding another temperature component,  the discrepancy of Fe-K lines
become smaller (Figure \ref{mos_apec}).
However, the $\chi^2$ is still worse than the single temperature MEKAL
model, and the radio of S lines cannot be well fitted.

In summary, the changes in the version 1.0 of the APEC code for the
Ly$\alpha$ lines (H-like ions) are giving a much less consistent picture
than the results from the MEKAL code.
Therefore, we are not sure that the new atomic data included in
APEC constitute an improvement.

\begin{figure}[]
\resizebox{\hsize}{!}{\includegraphics{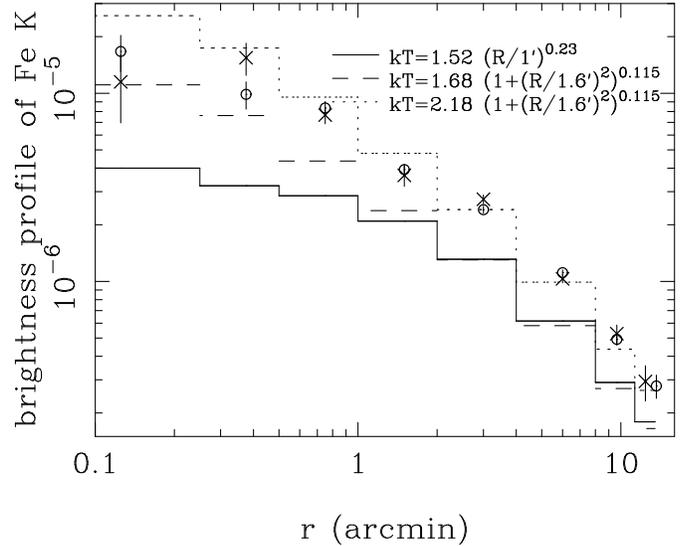}}
\caption{
Projected line brightness profile of Fe-K of the EMOS (open circles) and
 EPN (crosses). Errors correspond to 68\% confidence level.
Solid line and dashed lines represent the results
  using APEC model and deprojected temperature profile,
 $kT_{\rm{whole}}$ and $kT_{\rm{hard}}$,
respectively.
    We also plotted the profile for temperatures
 shifted by 30\% (dotted line)
}
\label{fekapec}
\end{figure}

%
%
}

\section{Intrinsic absorption}

\begin{figure}[]
\resizebox{\hsize}{!}{\includegraphics{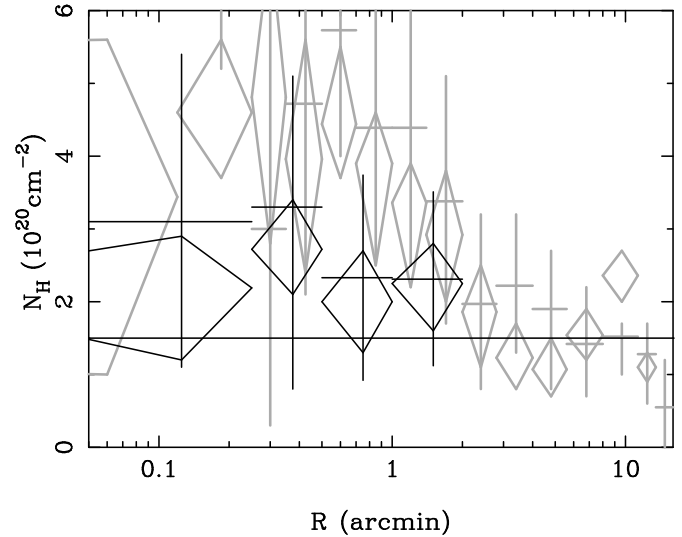}}
\caption{
The radial profile of hydrogen column density of EPN (crosses) and EMOS (diamonds) 
using the single temperature (gray) and the 
two temperature (black) MEKAL model fit.
Holizontal line corresponds to the column density  obtained from Nulsen \& B\"ohringer (1995).
}
\label{nh}
\end{figure}

%

The hydrogen column density, $N_{\rm H}$, derived from the single MEKAL fit
is almost constant at $R>2'$ (Table 2; Figure \ref{nh}).
 As summarized in table 2 and 5, the $N_{\rm H}$ derived from the single APEC model is consistent with
those from the single MEKAL model.
The average value at $R>2'$ is 1.5 (1.2-1.7)$\times10^{20}\rm{cm^{-2}}$ and
1.3 (1.0-1.5) $\times10^{20}\rm{cm^{-2}}$ for the EPN and EMOS, respectively.
Within 2 arcmin, the two phase MEKAL model fit gives a consistent value with those at $R>2'$,
although the single MEKAL model fit gives a higher $N_{\rm H}$ value by
several $\times10^{20}\rm{cm^{-2}}$.
The  response matrices of the EPN of September 2000 and November 2000 give
systematically different values by 1 $\sim2$ $\times10^{20}\rm{cm^{-2}}$.
Therefore, there may be a systematic uncertainty of  a few $ \times10^{20}\rm{cm^{-2}}$.

The value determined by the ROSAT PSPC observation is
about 1.5(1.4-2.0)$\times10^{20}\rm{cm^{-2}}$ (Nulsen and B\"ohringer 1995; B\"ohringer 1999).
The observed HI column density of our Galaxy by Stark et al. (1992) is 2.5$\times10^{20}\rm{cm^{-2}}$,
which is slightly reduced to $1.8 \times10^{20}\rm{cm^{-2}}$, by Lieu et al. (1996).
The observed $N_{\rm H}$ is consistent with the value obtained by ROSAT and also the Galactic column density.
In addition, the power-law component from the central AGN shows no significant excess absorption.
Thus, the intrinsic absorption should be negligible in the M~87 system.

\section{Density  profile}

\begin{figure}[]
\resizebox{\hsize}{!}{\includegraphics{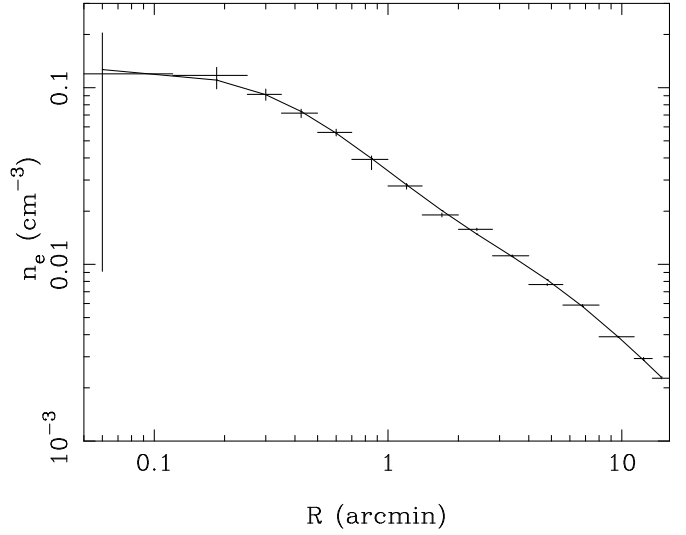}}
\caption{
Electron density profile.
The solid line is the best-fit double-$\beta$ model.
}
\label{ne}
\end{figure}

We have assumed $N_H$ to be the average value of those obtained at $R>2'$,
since the higher $N_H$ at $R<2'$ from the single MEKAL model may be artificial.
We have fixed the C, N, O abundance to $0.51 (1+(R/0.76')^2)^{-0.09} $, which is
the best fit $\beta$-model of the abundance profile of these elements.
Since the cooler component is distributed around the radio structures,  and 
the  hotter component is almost spherically symmetric,  the latter should dominate.
Thus, within 0.35 to 1.4$'$, we fitted the EPN spectra with a two
component MEKAL model and 
adopted the electron density of the hotter component.
The  electron density ($n_{\rm{e}}$), profile derived is shown in Table 6 and Figure \ref{ne}.
This results are  mostly consistent with those obtained by Nulsen and B\"ohringer (1995),
when considering the difference in the assumption of the abundance.

We  fitted the $n_{\rm{e}}^2$  profile with a sum of two $\beta$ models, 
a compact and an extended one, both centered on M~87.
We fixed the $\beta$ of the extended component to be 0.47.
which is the value obtained by ROSAT All Sky Survey (B\"ohringer et al. 1994).
The obtained best-fit parameters are shown in Table 7.
The two $\beta$ components have distinct parameters,
and cross over at $R \sim 2'$, or $\sim 10$ kpc.
\begin{table}[t]
 \begin{tabular}{rcrc}
\hline
$R$ & $n_e$ & $R$ & $n_e$ \\
(arcmin) & ($\rm{cm}^{-3}$) &(arcmin) & ($\rm{cm}^{-3}$)\\
\hline
0.00-0.12  &   1.2$^{+0.9}_{-1.1}\times10^{-1}$ & 2.00-2.80  &   1.58$^{+0.02}_{-0.02}\times10^{-2}$ \\ 
0.12-0.25  &   1.2$^{+0.1}_{-0.2}\times10^{-1}$	& 2.80-4.00  &   1.12$^{+0.01}_{-0.01}\times10^{-2}$ \\ 
0.25-0.35  &   9.2$^{+0.7}_{-0.7}\times10^{-2}$	& 4.00-5.60  &   7.69$^{+0.09}_{-0.09}\times10^{-3}$ \\ 
0.35-0.50  &   7.2$^{+0.4}_{-0.4}\times10^{-2}$	& 5.60-8.00  &   5.88$^{+0.05}_{-0.06}\times10^{-3}$ \\ 
0.50-0.70  &   5.6$^{+0.3}_{-0.3}\times10^{-2}$	& 8.00-11.3 &   3.89$^{+0.02}_{-0.03}\times10^{-3}$ \\ 
0.70-1.00  &   3.9$^{+0.2}_{-0.5}\times10^{-2}$	& 11.30-13.5&   2.93$^{+0.04}_{-0.04}\times10^{-3}$ \\  
1.00-1.40  &   2.8$^{+0.1}_{-0.1}\times10^{-2}$	& 13.50-16.0&   2.27$^{+0.02}_{-0.02}\times10^{-3}$ \\  
1.40-2.00  &   1.9$^{+0.1}_{-0.1}\times10^{-2}$\\
\hline
 \end{tabular}
\caption{Electron density profile}
\end{table}
\begin{table}[]
\begin{tabular}[t]{rrr}
\hline
          $R_{\rm{core}}$ & $\beta$ & ${n_e}_{\rm{center}}$\\
       (arcmin) & & (cm$^{-3}$)\\
\hline
  0.35 (0.27-0.44) & 0.42(0.39-0.46) & 0.13 (0.11-0.16)\\
   4.4 (4.2-4.7)  & 0.47 (fix) & 0.011 (0.010-0.012)\\
\hline
\end{tabular} 
\caption{result of double $\beta$ fit}
\end{table}

\section{Dark matter profile}

The total mass $M(R)$ within a 3-dimensional radius $R$, assuming 
hydrostatic equilibrium and  spherical symmetry, is given by
(e.g Sarazin 1988),
 $$M(R) = - \frac{kTR}{G \mu m_{\rm p}}
          \left( \frac{d \ln n}{d \ln R} + 
          \frac{d \ln T}{d \ln R} \right)   $$
where $m_{\rm p}$ is the proton mass, $k$ is the Boltzmann constant,
$G$ is the constant of gravity, and $\mu \sim 0.63$ is the mean
particle mass in units of $m_{\rm p}$.%

\begin{figure}[]
\resizebox{\hsize}{!}{\includegraphics{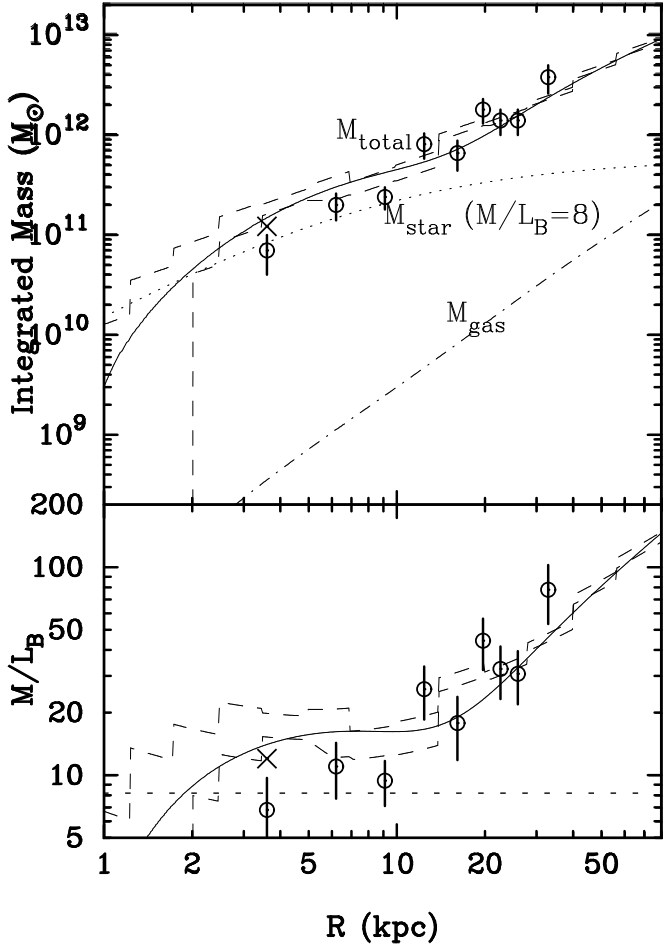}}
\caption{
(upper panel) Integrated mass profiles of the total gravitating matter (dashed lines for upper and lower limit
and solid line for the double $\beta$ model fit), 
the X-ray emitting plasma ( dot-dashed line), and the stellar component (dotted line)
assuming stellar $M/L_{\rm{B}}$ to be 8, in the M87~system.
The dynamical mass obtained by stars (cross; Kronawitter et al. 2000) and globular clusters
(open circles; Cohen \& Ryzhov 1997) are also shown.
(lower panel) Total mass-to-light ratio ($M/L_{\rm {B}}$) of the system.
The meanings of symbols and lines are the same as in the upper panel.
}
\label{m87mass}
\end{figure}

Based on the deprojected temperature and density profile obtained in \S3.2 and \S5,
we have calculated the gravitational mass of the M 87 system, together with the gas mass profile
using the power-law relation in Figure 3, and the double-$\beta$ relation  approximated as in Table 7.
The upper and lower limit of the mass profile is calculated considering
the uncertainty of the density gradient
and the temperature.  The temperature gradient is assumed to be smooth and expressed as the power-law relation.
 The upper and lower limit of the density gradient of the $i$-th shell is obtained
 from the  ratio of the value within $i+1$-th shell to that within  $i-1$-th shell.
As plotted in Figure \ref{m87mass},
our result is mostly consistent with the   dynamical mass profiles obtained by stars (Kronawitter et al. 2000) 
and globular clusters (Cohen \& Ryzhov 1997).

The calculated mass-to-blue light ratio (hereafter  $M/L_{\rm{B}}$)
of the stellar population ranges between 6 and 10 (e.g. Kronawitter et al. 2000).
Assuming a constant stellar  $M/L_{\rm{B}}$  of 8,
in Figure \ref{m87mass}, we also plotted the stellar mass profile, calculated from the
 luminosity profile in Giraud (1999).
Within nearly 0.5$r_e$, the stellar velocity dispersion profile indicates that the
dynamical $M/L_{\rm{B}}$ is $\sim$11 (e.g. Kronawitter et al. 2000).
Here, $r_e$ is the effective radius of M 87, which is 7.8 kpc.
Our mass profile indicates that at $R$=1-2$r_e$, the $M/L_{\rm{B}}$ is about 17 and beyond this radius,
the slope of the mass profile increases and $M/L_{\rm{B}}$ reaches $\sim 200$ at 10$r_e$.

\section{Application of the  cooling flow model}

\subsection{Mass flow profile from imaging data}

The steady state cooling flow structure in a spherically symmetric cluster is
obtained by,

$$\frac{1}{R^2}\frac{d}{dR}\{R^2\rho_gv\{\frac{v^2}{2}+\frac{5}{2}\frac{P}{\rho_g}+\phi(R)\}\}=
n_e^2\Lambda$$
$$\dot{M}=4\pi R^2\rho_g v$$
(e.g. Sarazin 1988),
where $\rho_g$, $v$, and $P$ are, respectively, the gas density, velocity, and pressure,
and $\phi(R)$ is the gravitational potential, $\Lambda$ is the emissivity.

\begin{figure}[]
\resizebox{\hsize}{!}{\includegraphics{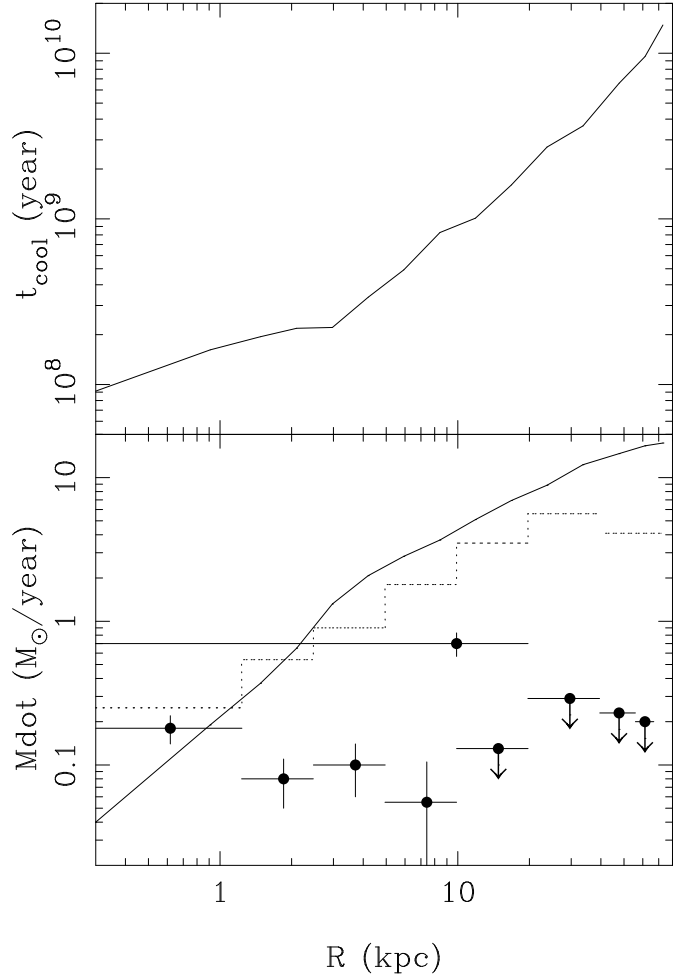}}
\caption{
Cooling time (upper panel) and  $\dot{M}_{\rm{I}}$ profile (lower panel).
Data corresponds to $\dot{M}_{\rm{S}}$, assuming cooling to 0 keV.
The dotted line is the differential of the  $\dot{M}_{\rm{I}}$, the mass deposition rate in each shell.
}
\label{cool}
\end{figure}

Even within the optical radii of elliptical galaxies, 
the X-ray luminosity is a factor of 20 larger 
than heating by stellar mass loss (Matsushita 2001). Therefore, we can neglect this component
as a sufficient heat source to balance cooling.
The cooling time is lower than the Hubble time within $R\sim$ 50 kpc (Figure \ref{cool}).
As shown in Fig. \ref{cool}, the  $\dot{M}_{\rm{I}}$ obtained ~is roughly proportional  to the radius. 
At $R\sim$ 50 kpc,  $\dot{M}_{\rm{I}}$ ~is 10 $M_\odot$,
which  is consistent to the classical value obtained by
Stewart et al. (1984) from the Einstein observation.

\subsection{Spectral fitting using a spectral cooling flow model}

With the RGS we can observe only the very central core of cooling flow clusters.
In contrast, using the EPN, we can accumulate spectra within the entire cooling flow region.

At $R=4'$, the derived cooling time  is only a few Gyr and $\dot{M}_{\rm{I}}$ is  8 $M_\odot \rm{y^{-1}}$, respectively.
The net effect of resonance scattering is negligible for the radiative output in this 
integrated circular region.
Without any heating, the  gas should cool down to 0 keV, emitting X-rays at intermediate temperatures.
Therefore, the spectral cooling rate, $\dot{M}_{\rm{S}}$, within the radius should be $8 M_\odot \rm{y^{-1}}$.

\begin{figure}[]
\resizebox{\hsize}{!}{\includegraphics{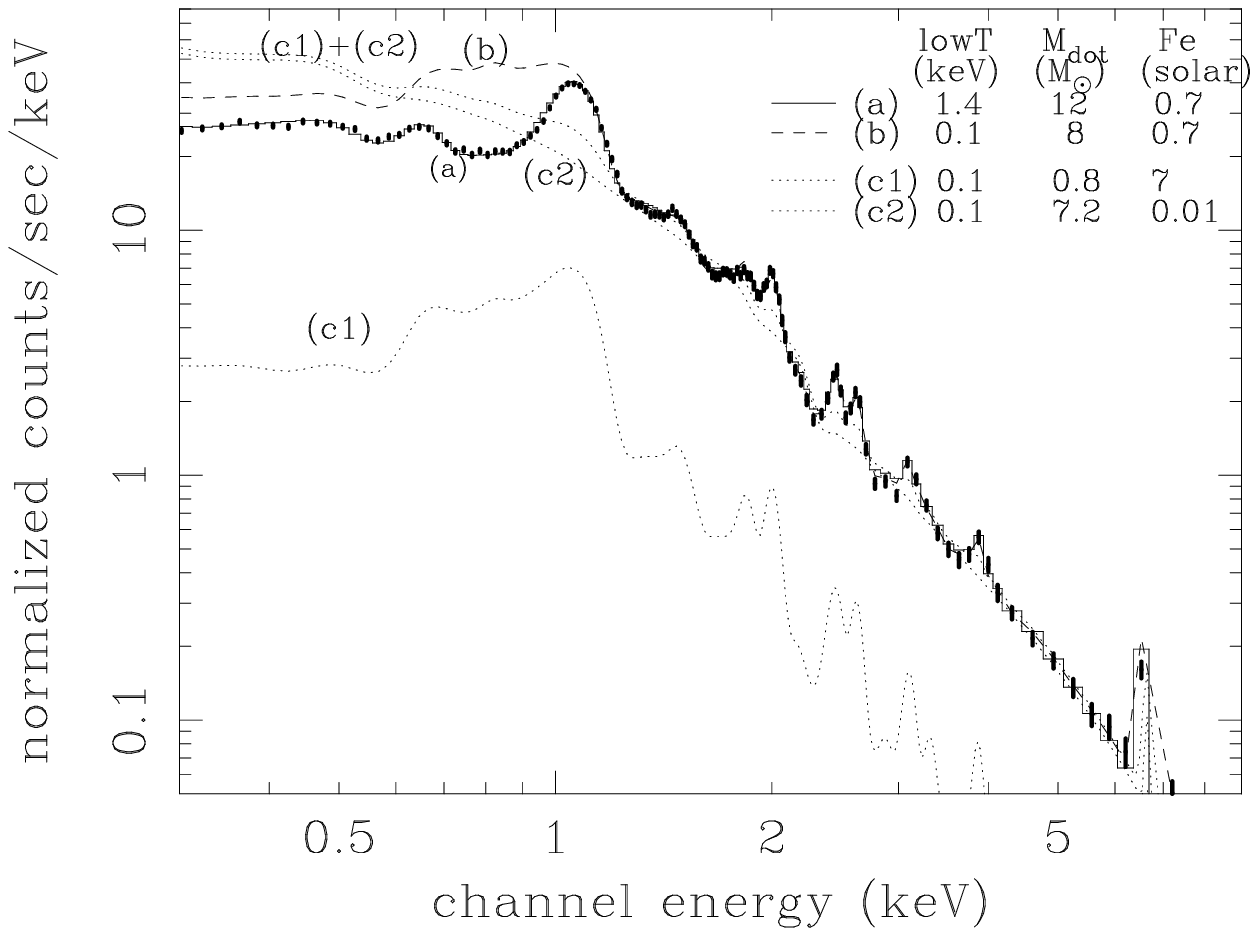}}
\resizebox{\hsize}{!}{\includegraphics{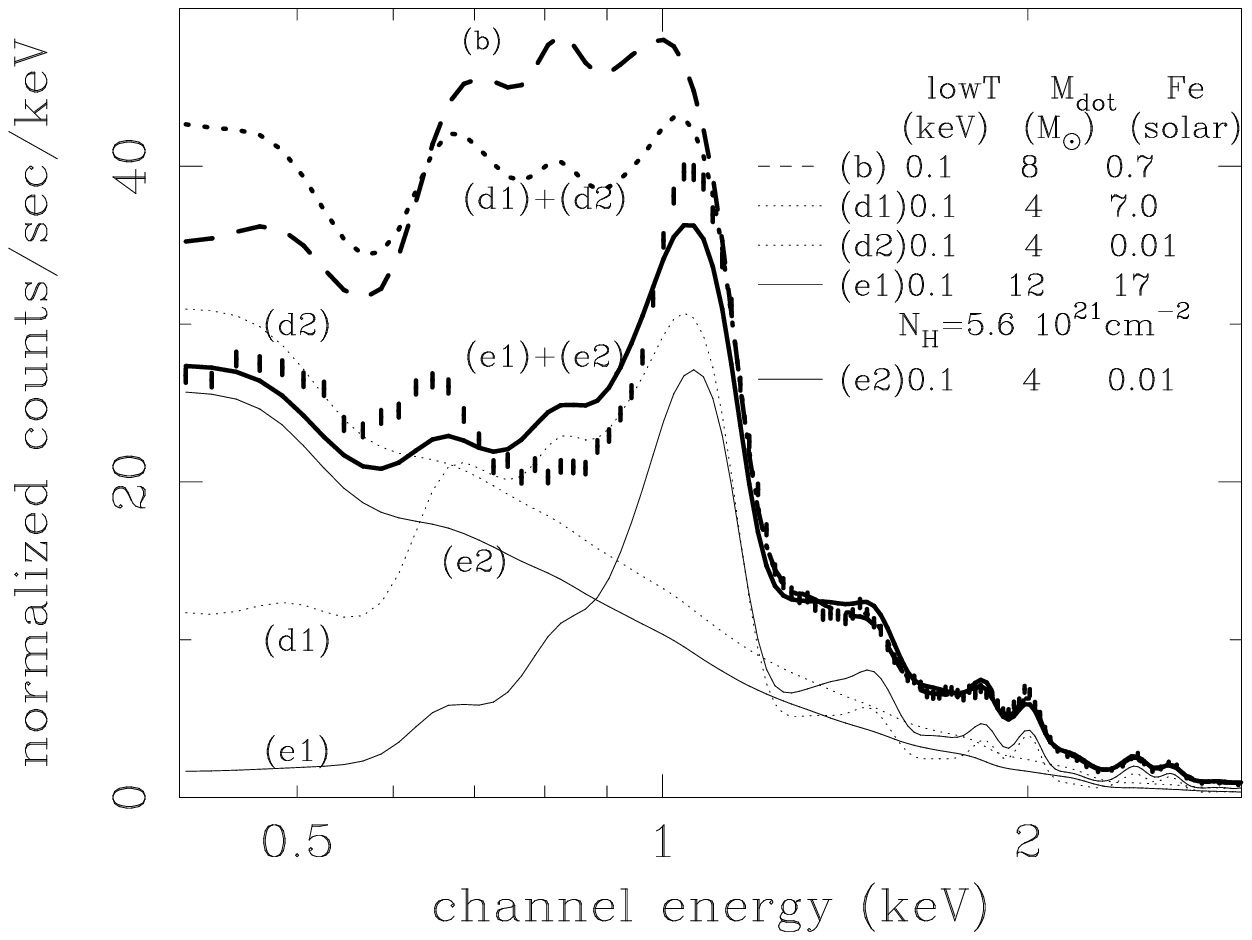}}
\caption{
 Deprojected spectra within $R<4'$ fitted with  a cooling flow model
with a cut-off temperature, lowT=1.44 keV (solid line, a).
The bold dotted line (b) shows the model with cooling to 0.1 keV.
The thin dashed lines of the upper panel, (c1) and  (c2), corresponds to the model with 
a model with 10 times higher metallicity and a model with low metal
 abundance,  respectively. The spectral cooling flow rates are
 summarized in the panel. The sum of (c1) and (c2) is also plotted.
The dotted lines of the lower panel, (d1) and (d2) are components of another bimodal
 metal abundance mdoel, and the solid lines of (e1) and (e2) corresponds
 to the model with excess absorption applied for the metal rich components.
}
\label{spec_cool}
\end{figure}

We  fitted the deprojected spectra within $R<4'$ with a cooling flow
model using the MEKAL code modified by photoelectric absorption.
As in other cooling flow clusters observed by RGS,
in order to fit the spectra, we still need a cut-off temperature, lowT, which is found 
to be 1.4 keV (Table 8).
Figure \ref{spec_cool} shows the EPN spectrum within $R<4'$, fitted with the cooling flow model.
We also plotted in the figure, 
another cooling flow model with lowT=0.1 keV, fitted only to the data in the spectral range  above 1.2 keV.
The difference in the continuum level between 0.2 and 0.5 keV
is due to the instrumental low-energy tail of Fe-L and O-K lines.
The  component  cooling radiatively to 0 keV should emit
strong Fe-L lines between 0.6 to 1.0 keV, which are not seen in the spectrum.

When adding an isothermal MEKAL component, we can fit the 
spectra with a cooling flow component with lowT=0.1 keV.
However,  $\dot{M}_{\rm{S}}$  is only 0.8 $M_\odot \rm{y^{-1}}$,
which is a factor of 10 smaller than   $\dot{M}_{\rm{I}}$ at the same radius.


\begin{table*}
 \begin{tabular}[t]{rccccccrr}
\hline                       
$R$  &model  &$kT_{\rm{vmekal}}$  &low$kT_{\rm{cf}}$   &
  high$kT_{\rm{cf}}$   & $N_{\rm{H}_{\rm{MEKAL}}}$ &
  $N_{\rm{H}_{\rm{MEKAL}}}$ & \mdot&$\chi^2/\mu$ \\
arcmin &    & (keV) & (keV) & (keV) & ($10^{20}\rm{cm^{-2}}$) & ($10^{20}\rm{cm^{-2}}$) & ($M_\odot$) & \\
\hline
0.0-4.0& mkcflow&--- &1.442$^{+0.002}_{-0.002}$ & 2.30$^{+0.02}_{-0.02}$
  & 1.3$^{+0.2}_{-0.2}$ & =$N_{\rm{H}_{\rm{MEKAL}}}$ & 13.8$^{+0.01}_{-0.01}$ & 370/191 \\
0.0-4.0& mekal+mkcflow &1.75$^{+0.01}_{-0.01}$& 0.10 (fix) &
  =$kT_{\rm{vmekal}}$ &2.9$^{+0.2}_{-0.2}$ & =$N_{\rm{H}_{\rm{MEKAL}}}$ & 0.63$^{+0.20}_{-0.08}$
  &416/130\\
0.0-4.0& mekal+mkcflow &1.58    & 0.10 (fix) & =$kT_{\rm{vmekal}}$ & 1.5
  &117& 8.0 (fix)  &1088/192\\
\hline
0.0-0.25 & mekal+mkcflow &1.7$^{+0.8}_{-0.6}$& 0.10 (fix) &
  =$kT_{\rm{vmekal}}$ &1.5 (fix) & 1.5 (fix) &0.18$^{+0.03}_{-0.04}$  &30/51\\
0.25-0.5 &mekal+mkcflow &1.58$^{+0.08}_{-0.08}$& 0.10 (fix) &
  =$kT_{\rm{vmekal}}$ &2.5$^{+0.6}_{-0.5}$ & =$N_{\rm{H}_{\rm{MEKAL}}}$ & 0.08$^{+0.03}_{-0.02}$  &62/51\\
0.5-1.0 &mekal+mkcflow &1.52$^{+0.05}_{-0.06}$& 0.10 (fix) &
  =$kT_{\rm{vmekal}}$ &4.4$^{+1.0}_{-1.0}$ &=$N_{\rm{H}_{\rm{MEKAL}}}$& 0.10$^{+0.04}_{-0.03}$  &91/51\\
1.0-2.0 &mekal+mkcflow &1.65$^{+0.06}_{-0.06}$& 0.10 (fix) &
  =$kT_{\rm{vmekal}}$ &2.8$^{+1.3}_{-1.3}$ &=$N_{\rm{H}_{\rm{MEKAL}}}$& 0.05$^{+0.05}_{-0.04}$  &78/51\\
2.0-4.0 &mekal+mkcflow &1.97$^{+0.05}_{-0.05}$& 0.10 (fix) &
  =$kT_{\rm{vmekal}}$ &1.5$^{+0.5}_{-0.5}$ &=$N_{\rm{H}_{\rm{MEKAL}}}$ & $<0.13$  &103/51\\
4.0-8.0 &mekal+mkcflow &2.30$^{+0.07}_{-0.08}$& 0.10 (fix) &
  =$kT_{\rm{vmekal}}$ &1.6$^{+0.4}_{-0.4}$&=$N_{\rm{H}_{\rm{MEKAL}}}$ & $<0.29$ &58/51\\
8.0-11.3&mekal+mkcflow &2.60$^{+0.12}_{-0.13}$& 0.10 (fix) &
  =$kT_{\rm{vmekal}}$ &  1.8$^{+0.8}_{-0.9}$  & =$N_{\rm{H}_{\rm{MEKAL}}}$& $<0.23$  &51/51
\\
11.3-13.5&mekal+mkcflow &2.66$^{+0.16}_{-0.16}$& 0.10 (fix) &
  =$kT_{\rm{vmekal}}$ & 1.5$^{+0.6}_{-0.6}$&=$N_{\rm{H}_{\rm{MEKAL}}}$ & $<0.20$  &111/51\\
\hline
 \end{tabular}
\caption{Result of spectral fitting using a spectral cooling flow model}
\end{table*}

\subsubsection{Test of a bimodal abundance model}

Fabian et al. (2001) made a proposal  to solve the cooling flow
problem by a scenario in which metals in the ICM are not uniformly distributed.
In this suggestion, a metal poor part of the gas cools without emitting lines and
a metal rich part cools rapidly.  In this way, the total  strength of Fe-L line emission should be reduced.
A bimodal abundance model, with 0.01 solar abundances and 10 times higher abundances than the
single abundance model and   mass deposition rates of 7.3 and 0.7
$M_{\odot}/yr$, respectively,  is shown in Figure \ref{spec_cool}.
It is very different from the observed spectrum in the Fe-L structure and
strength of K-$\alpha$ lines (Figure \ref{spec_cool}), 
since it reduces not only Fe-L lines but also K lines of Si, S, Ar,
Ca, and Fe (see also B\"ohringer et al. 2002).
In order to obtain the observed strength of K lines with these mass
deposition rates, the abundances in the high metallicity component should be $\sim$ 100 solar.
Therefore, we have tried another bimodal metal abundance model, 
where the mass deposition rates of the two components are allowed  to be free parameters, and
 fitted the spectra above 1 keV.
Figure  \ref{spec_cool} also shows the best fit model, extrapolated to lower energies.
With the mass deposition rates of 4 and 4 $M_{\odot}/yr$, for the metal
rich and metal poor components respectively, the spectra above $\sim$ 1
keV can be fitted.
Because of the dependence of the cooling function on metallicity, 
the Fe-L strength below 0.9 keV is slightly reduced compared to that
of the single abundance model.
However, the model shows very strong Fe-L lines below 0.9 keV.
We then added absorption for the metal rich components and then allowed 
the abundances of the metal rich component to vary.
Although including a column density of  5$\times10^{21}\rm{cm^{-2}}$ 
dramatically improved the fit,  the fit is much worse compared to the
fit with the cut-off temperature.
In addition,  such high absorption heavily reduces the spectra around
the 0.67 keV O line.
Increasing the O abundance of the metal rich components does not solve the problem, since it reduces
the strength of Fe-L lines due to the difference in the cooling function.
 Therefore, there must exist an unabsorbed component
whose O abundance is not so peculiar.
For intrinsic absorption,  more details are discussed also on
B\"ohringer et al. (2002).

In summary, any multi-abundance model cannot explain the observed Fe-L profiles.
In order to explain it, we need a sharp cut off in the temperature distribution.
%

\subsubsection{Radial profile of spectral cooling flow component}

The observed radial profile of $\dot{M}_{\rm{I}}$
is not constant, but roughly proportional to $R$.
The differential of $\dot{M}_{\rm{I}}$ between a shell and next shell
implies that we should observe a cooling flow component in the spectrum of each radial shell.
We thus fitted deprojected spectra with a 
cooling flow model with lowT=0.1 keV, and an isothermal MEKAL model.
The result is summarized in Table 8 and Figure \ref{cool}.
Within $R<1'$, the fit has improved from the single temperature fit.
However,  the $\dot{M}_{\rm{S}}$ values,   are a factor of 10 to 20 lower than 
the differential of $\dot{M}_{\rm{I}}$.
Only for the innermost shell, $\dot{M}_{\rm{S}}$ and $\dot{M}_{\rm{I}}$ are consistent with each other.

\section{Discussion}


\subsection{Temperature structure}

The deprojected spectra of M 87 strongly indicate that excluding the
region associated to the radio structures, the ICM is  single phase;
at a given radius, the ICM is dominated by  one temperature component.
We cannot detect any multi-phase cooling flow component  as seen by Allen et al. (2001).
The ICM is also not characterized by a two-phase structure as suggested by ASCA observations of the Centaurus cluster
(Fukazawa et al. 1994; Ikebe et al. 1999), and M~87 (Matsumoto et al. 1996).

 When excluding the regions with the radio structures, the
ICM is nearly spherically symmetric.
Using the MEKAL model,
a single phase  model can fit the deprojected spectra well at $R>2'$.
The temperature profile has a positive temperature gradient, $\sim 1.7$ keV at $R=1$'
and $\sim$ 2.5 keV at $R$=15$'$.
The consistency of the temperatures obtained by the continuum, Fe-L
region, Si, and S line ratios,  Fe-L to Fe-K ratio
indicate that the ICM is dominated by a single temperature component.
The regions associated to the radio structures have an additional
temperature components of 1 keV.
At $R<2'$, since it is
difficult to filter the complicated radio structures within 2$'$ with the XMM
spatial resolution,  we observe the major ICM components with the
temperature of 1.7 keV, and  small amount of the 1 keV components.
The spatial distribution of the latter component indicates that
the cooler component also relates to the region corresponding to the radio
lobes (see also Belsole et al. 2001).


The R-S model gives a  temperature lower by 20\% than MEKAL model.
In order to fit the spectra, the R-S model requires an additional high temperature component.
However, the two temperature R-S model cannot explain the observed Si and S line ratios.

 Except for the Fe-K line strength, the APEC temperatures are consistent with the
MEKAL models within 10-15\%.
Due to changes of the strength of K$\alpha$ lines,  
the single phase APEC model cannot explain the strength of Fe-K lines,
although it can fit the Fe-L and continuum simultaneously. The S line
ratio is discrepancy by a factor of 1.5, but it introduces only a 10\% difference
in the temperature structure.

Resonant  scattering of line emission should be important in dense core of galaxy clusters.
(Gil'fanov et al. 1987, Tawara et al. 1997).
Shigeyama (1998) calculated the effect on M~87 system, and found that surface brightness profile
considerably decreased within 1-2' because many Fe-L lines are optically thick.
In M87, B\"ohringer et al. (2001) suggested that the resonant scattering may
explain the observed sharp abundance drop at center.
This effect may also change the spectra and the Fe-L line profile in the central region.
Part of the 1 keV component may be due to this effect.
Therefore, in order to  study the  temperature structure  within 1-2' arcmin exactly, we have to
consider  effect of the resonance scattering.
However, since the temperature obtained from the continuum spectra is close to that
from the whole energy band, the main temperature component should be
similar to the obtained values.
 At least using the MEKAL model, the strength of the Fe-L and Fe-K are
consistent with each other, which have a different dependence on the
resonance line scattering.  
In addition,  most of the 1 keV component should be related to the radio
activity which has a  complicated spatial distribution (Belsole et al. 2001).
Most of the resonantly scattered light will be reemitted at slightly larger 
projected radii. Therefore integrating over the region, $R<4'$ will reduce the net effect
of scattering and the spectra will hardly be disturbed.

Matsushita (2001) found that normal elliptical galaxies 
have $\beta_{\rm{spec}}$=1.
The central stellar velocity dispersion of M~87 is 350 km/s (Whitemore et al. 1985).
The assumption of $\beta_{\rm{spec}}$=1 implies a temperature of 0.8 keV, which agrees 
well with the observed central ICM temperature of $M~87$.
The ICM temperature at 10--15$'$ of 2.5 keV  also agrees with the 
velocity dispersion of early-type galaxies of the Virgo cluster, which is 570 km/s (Binggeli and Tammann 1987).
The observed constancy in the temperature suggests that the central cool
temperature  mainly reflects the galaxy potential, 
even though some part of the cool component may be produced by a small cooling flow in the central region.




\subsection{Intrinsic absorption}

The hydrogen column density is almost flat within the field of view.
In addition, the spectrum of the central AGN does not show any absorption feature.
Therefore, intrinsic absorption should be negligible in the M 87 system.
If there is excess $N_H$ of the order of $10^{20}\rm{cm^{-2}}$ and the absorbing matter is uniformly distributed
within 10 kpc and 50 kpc, the mass of the absorbing matters is $10^8 M_\odot$ and $10^9 M_\odot$, respectively,
which is much smaller than the cooling flow  mass deposition rate accumulated over a Hubble time.

\subsection{Potential structure of ellipticals and clusters of galaxies}

We have determined the gravitational mass profile of the M 87 system, directly from the
observed density and temperature profiles.
At 0.5--2 $r_e$, $M/L_{\rm B}$ of the system is constant at $\sim17$.
Beyond this radius, it starts to increase to $\sim 200$ at 80 kpc.

\begin{figure}[]
\resizebox{\hsize}{!}{\includegraphics{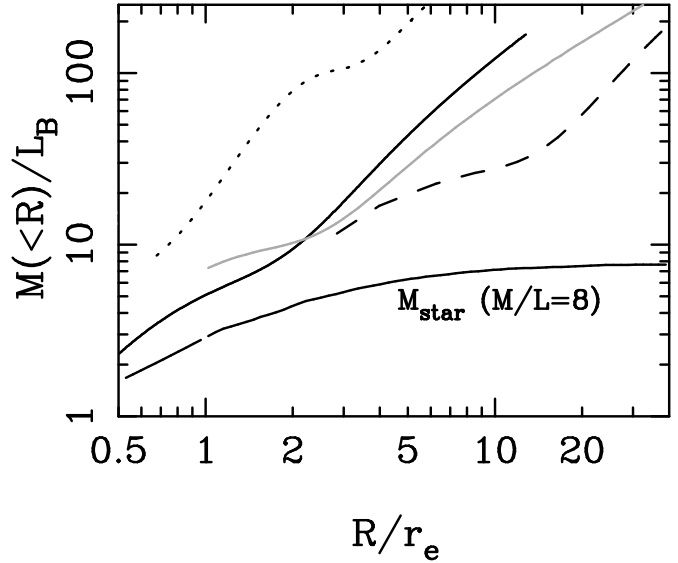}}
\caption{
Integrated mass profiles of the total gravitating matter scaled by
$L_B$, and the stellar component (solid line) in the M~87 (solid line; $L_X/L_{\sigma}$=22), the
Centaurus cluster (dotted line; Ikebe et al. 1999; $L_X/L_{\sigma}$=47), NGC 1399 (dashed line; Ikebe et al. 1996 ;$L_X/L_{\sigma}$=3),
and NGC 4636 (gray line; Matsushita et al. 1998; $L_X/L_{\sigma}$=12).
$L_X$ and $L_{\sigma}$ means the ISM luminosity within 4$r_e$ and
kinematical energy input from stellar mass loss (Matsushita 2001).
}
\label{mass_all}
\end{figure}

This structure is essentially the same feature as seen in the Fornax
cluster (Ikebe et al. 1996), the Centaurus cluster (Ikebe et al. 1999)
 and NGC 4636 (Matsushita et al. 1998).
Figure \ref{mass_all} summarizes the gravitational mass profiles of the 3 cD galaxies and NGC 4636.
These galaxies have similar hierarchical mass profiles;
the gravitational mass gradually increases up to several $r_e$,  where the
gravitational to stellar mass ratio is common among galaxies.
Then,  $M/L$ start to increase steeply.

Before the total to stellar mass ratio increases steeply, 
the total $M/L_{\rm {B}}$ is  $\sim 17$ at 1 $r_e$, and $\sim30$ even at 10 $r_e$.
This feature should reflect the potential of elliptical galaxies.
The  $M/L_{\rm {B}}$ of the central core of these galaxies obtained by
the dynamics and the same ratio obtained by the stellar population
synthesis models are consistent with each other giving a typical value
of 6--10 (e.g. Kronawitter et al. 2000).
Considering the observed metallicity gradient of stars,  the stellar  $M/L_{\rm {B}}$ should be slightly 
smaller in the outer regions (e.g. Kodama \& Arimoto 1997). Our result suggests that
the dark mass distribution of elliptical galaxies is slightly more extended than those of stars.
It is thought that elliptical galaxies have a 10 times larger amount of dark mass
than stellar mass (e.g. Ciotti et al. 1991).
However, the galaxies themselves may not  contain such large amount of  mass.
This result should be important for the study of the origin of the dark matter content in elliptical galaxies
and also for the study of the formation and evolution of these galaxies.

The scatter of the X-ray to blue luminosity ratio  of early-type galaxies has long been a problem
(e.g. Canizares et al. 1987).
Matsushita (2001) found that $L_X$ of most of the early-type galaxies
is well explained by a kinematical heating of the gas supplied by stellar mass loss
($L_\sigma$).
All of the galaxies which have much larger $L_X$ than $L_\sigma$
have very extended X-ray emission.
Figure \ref{mass_all} shows that the  galaxies with a larger $L_X/L_{\sigma}$
 have smaller radii of the starting point of the strong increase of $M/L$, when scaled $r_e$.
In other words,  the X-ray luminosity of cD galaxies may be determined in relation to their potential structure,
as indicated by Matsushita (2001).

\subsection{Problem of standard cooling flow model}

The standard cooling flow model indicates that the ICM is multi-temperature on small scales,
since it is thought that mass is deposited within the whole cooling flow region.
The cooling matter should emit strong Fe-L lines below 0.9 keV.
We cannot detect a significant spectral cooling flow component from the whole field of view of the detector
and the ICM at any given radius is dominated by  a single temperature component.
The upper limit on $\dot{M}_{\rm{S}}$  is  an order of magnitude smaller than $\dot{M}_{\rm{I}}$.
A bimodal metal abundance distribution model cannot explain the observed Fe-L profile.

The resonance scattering may reduce the strength of some resonance lines.
However,   all the Fe-L lines below 0.9 keV,  which includes both resonance lines and non-resonance lines,
are suppressed compared with the expected value for a cooling flow.
In addition,  the spectral cooling flow component is also small where the resonance scattering
is not effective.
Therefore,  the resonant  scattering can   not be responsible for the
 suppresion of  the Fe-L lines below 0.9 keV.

At 1 $r_e$ (1.6$'$) of M 87,
$\dot{M}_{\rm{I}}$ is $\sim 4 M_\odot yr^{-1}$.
 In contrast,  the upper limit on $\dot{M}_{\rm{S}}$  is $\sim 0.4 M_\odot yr^{-1}$.
This value is close to the stellar mass loss rate within 1 $r_e$, which
is 0.5  $M_\odot yr^{-1}$. 
At this radius,  $t_{\rm cool}$ is $ 10^9 \rm{yr}$ and the gas mass is
$2 ~10^9 M_\odot$.
Therefore, considering the observed abundance gradient, the standard
cooling flow model indicate that the gas  should be diluted
by metal poor surrounding gas.
However, the observed Fe abundance of M 87 within 1$r_e$, $\sim 1$ solar,  is similar to those
 of normal X-ray luminous galaxies, which are also
 about $\sim 1$  solar (Matsushita et al. 1997; 2000).
Note that the observed  ISM abundance of NGC 4636 from the RGS spectra
 is 0.9 solar (Xu et al. 2001).
In these systems, the enrichment has to be compared with a mass flow rate of $\sim 1 M_\odot {\rm{yr^{-1}}}$
as implied by the standard cooling flow models for these galaxies.
For the case of NGC 4636, the mass flow rate is determined to be $\sim 1 M_\odot {\rm{yr^{-1}}}$
 (Bregman, Miller, \& Irwin 2001).
The effect of the dilution by ICM should be determined by a ratio of
 the mass flow rate and the optical luminosity since metals ejected from
 a galaxy should be proportional to the stellar luminosity.
Within 1$r_e$, the former rate of NGC 4636 is a factor of 10 smaller
 than that of M 87, while the latter  is only a factor of 2.5
 smaller.
In a much larger cooling flow  as exists in M 87 in which the Fe enrichment
is supplied by a stellar population comparable to that of normal
ellipticals,   the relative enrichment
will be less and the abundance gradient must be reduced in contrast to the observed data.
Therefore, the overall cooling flow expected from the classical galaxy cluster cooling flow model,
should not exist in M 87, although
a small scale cooling flow, with mass deposition rate  close to the stellar mass loss rate,
 may exists  in the central core of the classical cooling flow regions.

\section{Conclusion}
The major result of this paper is strong evidence that the ICM in the halo of M 87
is isothermal locally, except probably for the very central region, where more
than one temperature component is present.
This could be partly due to the interaction effects of the ICM with the jet and the radio lobes 
of M 87. This finding and the fact that no spectral signature of low temperature components
below temperatures of 0.8 keV are observed is in disagreement with the standard cooling flow model
which predicts a strongly multi-phase structure of the ICM and a distinct temperature distribution and
resulting spectrum with clear signature of low temperature components.
We have also shown that strong inhomogeneities in the metal distribution cannot resolve this problem.
Therefore we conclude that the scenario for the dense gas with short cooling time in the centers of 
clusters needs to be revised.
Probably heating processes that can substantially reduced the mass deposition have
to be reconsidered, like the energy input of the central AGN.
The central temperature closely reflects the gravitational potential depth of the central galaxy, 
rather than the  existence of a cooling flow.

\begin{acknowledgements}

 This work was supported by the Japan
 Society for the Promotion of Science (JSPS) through its Postdoctoral
 Fellowship for Research Abroad and Research Fellowships for Young
 Scientists.
\end{acknowledgements}



 \end{document}